\documentclass[a4paper,10pt]{article}
\usepackage[dvips]{graphicx}
\usepackage{amssymb,amsmath}
\numberwithin{equation}{section}

\begin{document}

\title{ Knot Universes in   Bianchi Type  I Cosmology}

\author{Ratbay Myrzakulov\\ \textit{Eurasian International Center for Theoretical Physics,  } \\ \textit{Eurasian National University, Astana 010008, Kazakhstan} }

\maketitle
\date{}

\begin{abstract}We investigate the trefoil and figure-eight knot universes from Bianchi-type I cosmology. 
In particular, we construct several concrete models describing the knot universes related to the cyclic universe and examine those cosmological features and properties in detail. Finally some examples of unknotted closed curves solutions (spiky and Mobius strip universes) are presented.
 
\end{abstract}
\date{}
\tableofcontents
%\newpage
\section{Introduction}

Inflation is one of the most important phenomena in modern cosmology 
and has been confirmed by recent observations on cosmic microwave background (CMB) radiation~\cite{WMAP}. 
Furthermore, it is suggested 
by the cosmological and astronomical observations of 
Type Ia Supernovae~\cite{SN1}, CMB radiation~\cite{WMAP}, 
large scale structure (LSS)~\cite{LSS}, 
baryon acoustic oscillations (BAO)~\cite{Eisenstein:2005su}, 
and weak lensing~\cite{Jain:2003tba} that 
the expansion of the current universe is accelerating. 
In order to explain the late time cosmic acceleration, we need to 
introduce so-called dark energy in the framework of general relativity 
or modify the gravitational theory, which can be regarded as a kind of geometrical dark energy (for reviews on dark energy, see, e.g.,~\cite{
Copeland:2006wr}-\cite{Li:2011sd}, and for reviews on modified gravity, 
see, e.g.,~\cite{Review-Nojiri-Odintsov}-\cite{Clifton:2011jh}). 

It is considered that there happened a Big Bang singularity 
in the early universe. 
In addition, at the dark energy dominated stage, 
the finite-time future singularities will occur  \cite{Big-Rip}-\cite{Future-singularity-MG}. 
There also exists the possibility that a Big Crunch singularity will happen. 
To avoid such cosmological singularities, 
there are various proposals such as 
the cyclic universe~\cite{Cyclic-universe}-\cite{Sahni:2012er} 
(in other approach of the cyclic universe, see \cite{Chung:2001ka}), 
the ekpyrotic scenario~\cite{Ekpyrotic-scenarios}, 
and the bouncing universe~\cite{Bouncing-cosmology}.  
%Novello:2008ra}. 

On the other hand, as a related theory to the cyclic universe, 
the trefoil and figure-eight knot universes have been explored in Ref.~\cite{Knot-universe}. 
In the homogeneous and isotropic 
Friedmann-Lema\^{i}tre-Robertson-Walker (FLRW) and 
the homogeneous and anisotropic 
Bianchi-type I cosmologies, 
the geometrical description of these knot theories corresponds to 
oscillating solutions of the gravitational field equations. Note that the terms "the trefoil knot universe" and "the figure-eight knot universe" were  introduced for the first time in Ref.~\cite{Knot-universe}. 
Moreover, the Weierstrass $\wp(t)$, $\zeta(t)$ and $\sigma(t)$ functions and 
the Jacobian elliptic functions have been applied to solve several issues on 
astrophysics and cosmology~\cite{Elliptic-functions}. 
In particular, very recently, by combining the reconstruction method in Refs.~\cite{Review-Nojiri-Odintsov, Nojiri:2005sx, Nojiri:2005sr, Stefancic:2004kb} 
with the Weierstrass and Jacobian elliptic functions, 
the equation of state (EoS) for the cyclic universes~\cite{Bamba:2012gq} 
and periodic generalizations of Chaplygin gas type models~\cite{Kamenshchik}-\cite{Benaoum} for dark energy~\cite{JC2} have been examined. 
This procedure can be considered to a novel approach to cosmological models 
in order to investigate the properties of dark energy. 

In this paper, 
we explore the cosmological features and properties of the trefoil and figure-eight knot universes from Bianchi-type I cosmology in detail. 
In particular, we construct several concrete models describing the trefoil and figure-eight knot universes based on Bianchi-type I spacetime. 
In our previous work~\cite{Knot-universe}, the models of the knot universes 
from the homogeneous and isotropic FLRW spacetime were studied. By using the equivalent 
procedure, as continuous investigations, in this work we explicitly demonstrate that the knot universes can be constructed by Bianchi-type I spacetime. 
In other words, our purpose is to establish the formalism which can describe 
the knot universes. 

It is significant to emphasize that 
according to the recent cosmological data analysis~\cite{WMAP}, 
it is implied that the universe is homogeneous and isotropic. In fact, however, recently the feature of anisotropy of cosmological phenomena such as anisotropic inflation~\cite{anisotropic-inflation} 
has also been studied in the literature. In such a cosmological sense, 
it can be regarded as reasonable to consider the anisotropic universe 
including Bianchi-type I spacetime. 
%%%%% Units %%%%%
The units of the gravitational constant 
$8 \pi G = c =1$ with $G$ and $c$ being the gravitational constant and the 
seed of light are used.

The organization of the paper is as follows. 
In Sec.\ II, we explain the model and derive the basic equations. 
In Sec.\ III, we investigate the trefoil knot universe. 
Next, we study the figure-eight knot universe in Sec.\ IV.  In Sec. V we present some unknotted closed curve solutions of the model. 
Finally, we give conclusions in Sec.\ VI. 
%%%%%

\section{The model}
In this section we briefly review some basic facts about the
Einstein's field equation. We start from the standard gravitational
action (chosen units are $c=8\pi G=1$)
\begin{equation}
S=\frac{1}{4}\int d^{4}x\sqrt{-g}(R-2\Lambda+L_m),
\end{equation}
where $R$ is the Ricci scalar, $\Lambda$ is the cosmological
constant and $L_m$ is the matter Lagrangian. For a general metric
$g_{\mu\nu}$, the line element is
 \begin{equation}
ds^2=g_{\mu\nu}dx^\mu dx^\nu, \ \ (\mu,\nu=0,1,2,3).
\end{equation}
The corresponding Einstein field equations are given by
 \begin{equation}\label{KN2.3}
R_{\mu\nu}+\Big(\Lambda-\frac{1}{2}R\Big)g_{\mu\nu}=- \kappa^2T_{\mu\nu},
\end{equation}
where $R_{\mu\nu}$ is the Ricci tensor. This equation forms the
mathematical basis of the theory of general relativity. In \eqref{KN2.3},
$T_{\mu\nu}$ is the energy-momentum tensor of the matter field
defined as
  \begin{equation}
T_{\mu\nu}=\frac{2}{\sqrt{-g}}\frac{\delta L_m}{\delta g^{\mu\nu}},
\end{equation}
and satisfies the conservation equation
 \begin{equation}\label{KN2.5}
\nabla_\mu T^{\mu\nu}=0,
\end{equation}
where $\nabla_\mu$ is the covariant derivative which is the relevant
operator to smooth a tensor on a differentiable manifold. Eq.\eqref{KN2.5}
yields the conservations of energy and momentums, corresponding to
the independent variables involved. The general Einstein equation
\eqref{KN2.3} is a set of non-linear partial differential equations. We
consider the Bianchi - I  metric 
\begin{equation}\label{KN2.6}
ds^2=-d\tau^2+A^2dx_1^2+B^2dx_2^2+C^2dx_3^2,
\end{equation}
where we assume that $\tau=t/t_0, x_i=x_i^{\prime}/x_{i0}, A, B, C$ are dimensionless (usually we put $t_0=x_{i0}=1$).
Here the metric potentials $A,B$ and $C$ are functions of $\tau=t$ alone. This insures that the model is spatially homogeneous. 
The statistical volume for the anisotropic Bianchi type-I  model can be written as 
\begin{equation}
V=ABC.
\end{equation}
The Ricci scalar is 
\begin{equation}\label{KN2.8}
R=g^{ij}R_{ij}=2\left(\frac{\ddot{A}}{A}+\frac{\ddot{B}}{B}+\frac{\ddot{C}}{C}+\frac{\dot{A}\dot{B}}{AB}+\frac{\dot{A}\dot{C}}{AC}+\frac{\dot{B}\dot{C}}{BC}\right),
\end{equation}
where $\dot{A}=dA/d\tau$ and so on. The non-vanishing  components of Einstein tensor
\begin{equation}
G_{ij}=R_{ij}-0.5g_{ij}R
\end{equation}
are
\begin{eqnarray}
G_{00}&=&\frac{\dot{A}\dot{B}}{AB}+\frac{\dot{A}\dot{C}}{AC}+\frac{\dot{B}\dot{C}}{BC},\\
G_{AA}&=&-A^2\left(\frac{\ddot{B}}{B}+\frac{\ddot{C}}{C}+\frac{\dot{B}\dot{C}}{BC}\right),\\
G_{BB}&=&-B^2\left(\frac{\ddot{A}}{A}+\frac{\ddot{C}}{C}+\frac{\dot{A}\dot{C}}{AC}\right),\\
G_{CC}&=&-C^2\left(\frac{\ddot{B}}{B}+\frac{\ddot{A}}{A}+\frac{\dot{B}\dot{A}}{BA}\right).
\end{eqnarray}
We define $a=(ABC)^{\frac{1}{3}}$ as the average scale factor so that the average Hubble parameter  may be defined as
\begin{equation}
H=\frac{\dot{a}}{a}=\frac{1}{3}\left(\frac{\dot{A}}{A}+\frac{\dot{B}}{B}+\frac{\dot{C}}{C}\right).
\end{equation}
We write this average  Hubble parameter $H$ sometimes as 
\begin{equation}
H=\frac{1}{3}\left(H_{1}+H_{2}+H_{3}\right),
\end{equation}
where 
\begin{equation}\label{KN2.16}
H_{1}=\frac{\dot{A}}{A},\quad H_{2}=\frac{\dot{B}}{B},\quad H_{3}=\frac{\dot{C}}{C}\end{equation} are the directional Hubble parameters in the directions of $x_1, x_2$ and $x_3$ respectively. Hence we get the important relations
\begin{equation}\label{KN2.17}
A=A_0e^{\int H_{1}dt},\quad B=B_0e^{\int H_{2}dt},\quad C=C_0e^{\int H_{3}dt},\end{equation} 
where $A_0, B_0, C_0$ are integration constants.  
The other  important cosmological  quantity is the deceleration parameter $q$, which for our model reads as
\begin{equation}
q=-\frac{a\ddot{a}}{\dot{a}^{2}}.
\end{equation}

Next, we assume that  the energy-momentum tensor of fluid has the form 
\begin{equation}
T_{ij}=diag[T_{00},T_{11},T_{22},T_{33}]=diag[\rho,-p_{1},-p_{2},-p_{3}].
\end{equation}
Here $p_{i}$ are the pressures along the $x_i$ axes recpectively, $\rho$ is the proper density of energy.
Then the Einstein   equations (with gravitational units, $8\pi G=1$ and $c=1$) read as 
\begin{equation}
R_{ij}-\frac{1}{2}Rg_{ij}=-T_{ij},
\end{equation}
where we assumed $\Lambda=0$. For the metric \eqref{KN2.6} these equations take the form
\begin{eqnarray}\label{KN2.21}
\frac{\dot{A}\dot{B}}{AB}+\frac{\dot{B}\dot{C}}{BC}+\frac{\dot{C}\dot{A}}{CA}-\rho&=&0,\\
\frac{\ddot{B}}{B}+\frac{\ddot{C}}{C}+\frac{\dot{B}\dot{C}}{BC}+p_1&=&0,\\
\frac{\ddot{C}}{C}+\frac{\ddot{A}}{A}+\frac{\dot{C}\dot{A}}{CA}+p_2&=&0,\\
\frac{\ddot{A}}{A}+\frac{\ddot{B}}{B}+\frac{\dot{A}\dot{B}}{AB}+p_3&=&0.\label{KN2.24}
\end{eqnarray}
In terms of the Hubble parameters this system takes the form
\begin{eqnarray}\label{KN2.25}
H_1H_2+H_2H_3+H_1H_3-\rho&=&0,\\
\dot{H}_2+
\dot{H}_3+H^2_2+H^2_3+H_2H_3+p_1&=&0,\\
\dot{H}_3+\dot{H}_1+H^2_3+H^2_1+H_3H_1+p_2&=&0,\\
\dot{H}_1+
\dot{H}_2+H^2_1+H^2_2+H_1H_2+p_3&=&0.\label{KN2.28}
\end{eqnarray}
Also we can introduce the three EoS parameters as
\begin{equation}
\omega_1=\frac{p_1}{\rho},\quad \omega_2=\frac{p_2}{\rho},\quad \omega_3=\frac{p_3}{\rho}
\end{equation}
and the deceleration parameters
\begin{equation}
q_1=-\frac{\ddot{A}A}{\dot{A}^2},\quad q_2=-\frac{\ddot{B}B}{\dot{B}^2},\quad q_3=-\frac{\ddot{C}C}{\dot{C}^2}.
\end{equation}
Finally we want present the equation
\begin{equation}
2\dot{H}+6H^2=\rho-p,
\end{equation}
where
\begin{equation}
p=\frac{p_1+p_2+p_3}{3}
\end{equation}
is the average pressure. Hence we can calculate the average papameter of the EoS as
\begin{equation}
\omega=\frac{p}{\rho}=\frac{\omega_1+\omega_2+\omega_3}{3}.
\end{equation}
Let us also we present the expression of $R$ in terms of $H_i$. From \eqref{KN2.8} and \eqref{KN2.16} follows
\begin{equation}
R=2\left(\dot{H}_{1}+\dot{H}_{2}+\dot{H}_{3}+H_1^2+H_2^2+H_3^2+H_1H_2+H_1H_3+H_2H_3)\right).
\end{equation}
Now we want present the knot and unknotted universe solutions of the system \eqref{KN2.21}--\eqref{KN2.24} or its equivalent \eqref{KN2.25}--\eqref{KN2.28}. Consider some examples.
\section{The trefoil knot universe}
Our aim in this section is to construct simplest examples of the  knot universes, namely, the trefoil knot universes. Consider examples.
\subsection{Example 1.} 
\begin{figure}[h]
	\centering
		\includegraphics{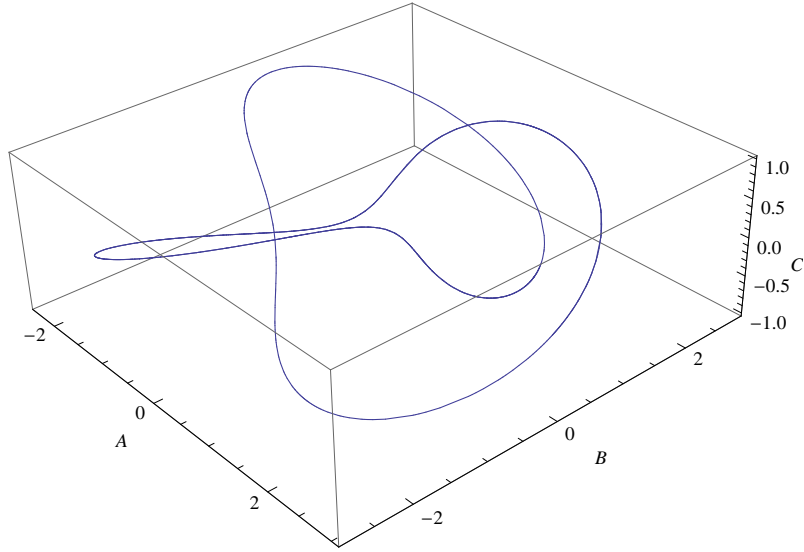}
	\caption{The trefoil knot for Eqs. \eqref{KN3.13}--\eqref{KN3.15}, where $A_0=B_0=C_0=0$.}
	\label{KN1}
\end{figure}

Let us  we assume that our universe is filled by the fluid with the following parametric EoS
\begin{eqnarray}
p_1&=&-\frac{D_1}{E_1},\\
p_2&=&-\frac{D_2}{E_2},\\
p_3&=&-\frac{D_3}{E_3},\\
\rho&=&\frac{D_0}{E_0},
\end{eqnarray}
where
\begin{eqnarray}
D_1&=&(-12\sin^2(3\tau)+36\cos(3\tau)+18\cos^2(3\tau))\cos(2\tau)-49\sin(2\tau)\notag\\
& &(26/49+\cos(3\tau))\sin(3\tau),\\
E_1&=&\sin(3\tau)(2+\cos(3\tau))\sin(2\tau),\\
D_2&=&-18\sin(2\tau)\cos^2(3\tau)+(-49\sin(3\tau)\cos(2\tau)-36\sin(2\tau))\cos(3\tau)-\notag\\
& &26\sin(3\tau)\cos(2\tau)+12\sin^2(3\tau)\sin(2\tau),\\
E_2&=&\sin(3\tau)(2+\cos(3\tau))\cos(2\tau),\\
D_3&=&-30\sin(3\tau)(2+\cos(3\tau))\cos^2(2\tau)-38\sin(2\tau)(\cos^2(3\tau)-(27/38)\sin^2(3\tau)+\notag\\
& &(58/19)\cos(3\tau)+40/19)\cos(2\tau)+30\sin(3\tau)\sin^2(2\tau)(2+\cos(3\tau)),\\
E_3&=&(2+\cos(3\tau))^2\cos(2\tau)\sin(2\tau),\\
D_0&=&(6\cos^2(2\tau)-6\sin^2(2\tau))\cos^3(3\tau)+(24\cos^2(2\tau)-22\sin(3\tau)\sin(2\tau)\cos(2\tau)-\notag\\
& &24\sin^2(2\tau))\cos^2(3\tau)+((-6\sin^2(3\tau)+24)\cos^2(2\tau)-52\sin(3\tau)\sin(2\tau)\notag\\
& &\cos(2\tau)+(6\sin^2(3\tau)-24)\sin^2(2\tau))\cos(3\tau)-(12(\cos(2\tau)-(3/4)\sin(3\tau)\notag\\
& &\sin(2\tau)))(\sin(3\tau)\cos(2\tau)+(4/3)\sin(2\tau))\sin(3\tau),\\
E_0&=&(2+\cos(3\tau))^2\cos(2\tau)\sin(2\tau)\sin(3\tau).
\end{eqnarray}
Substituting these expressions for the pressures and  density of energy into the system \eqref{KN2.21}--\eqref{KN2.24}, we obtain the following its solution 
\begin{eqnarray}\label{KN3.13}
A&=&A_0+[2+\cos (3\tau)]\cos (2\tau), \\
B&=&B_0+[2+\cos (3\tau)]\sin (2\tau), \\
C&=&C_0+\sin (3\tau),\label{KN3.15}
    \end{eqnarray}
 where $A_0, B_0, C_0$ are some real constants. We see that this solution describes the  trefoil knot. In fact the solution \eqref{KN3.13}--\eqref{KN3.15} is the parametric equation of the trefoil knot. In Fig. \ref{KN1} we plot the trefoil knot for Eqs. \eqref{KN3.13}--\eqref{KN3.15}, where we assume
 \begin{equation}\label{KN3.16}
A_0=B_0=C_0=0
\end{equation} and the initial conditions are $A(0)=3, B(0)=C(0)=0$. 
The Hubble parameters for the solution \eqref{KN3.13}--\eqref{KN3.15} with \eqref{KN3.16} read as
\begin{eqnarray}\label{KN3.17}
H_1&=&-2\tan(2\tau)-\frac{2\sin(3\tau)}{2+\cos (3\tau)}, \\
H_2&=&-2\cot(2\tau)-\frac{2\sin(3\tau)}{2+\cos (3\tau)}, \\
H_3&=&3\cot(3\tau).\label{KN3.19}
    \end{eqnarray}
  \begin{figure}[h]
	\centering
		\includegraphics{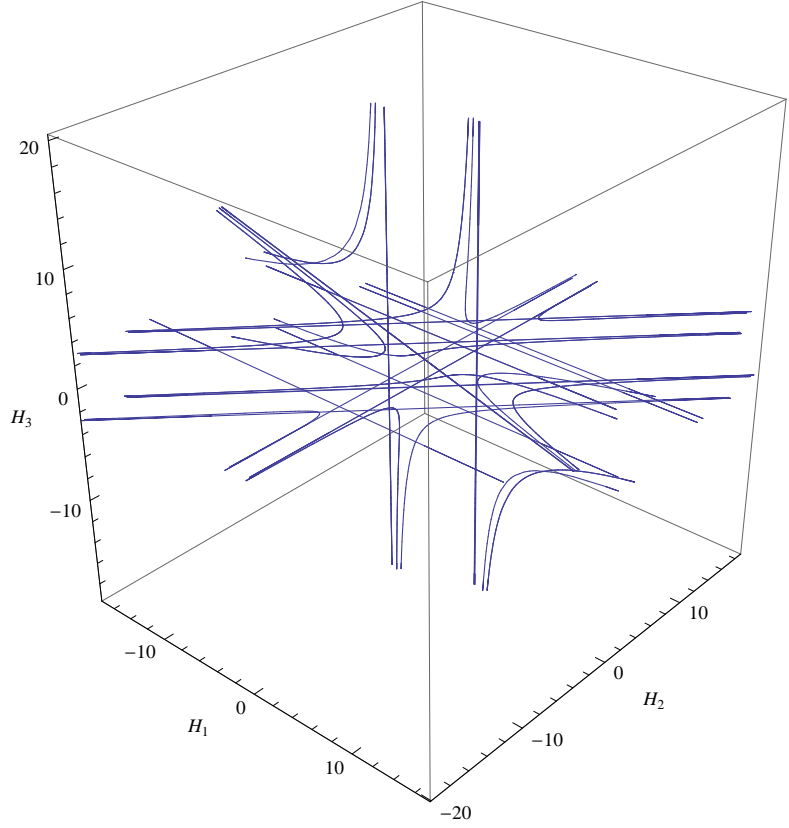}
	\caption{The  evolution of the Hubble parameters for Eqs. \eqref{KN3.17}--\eqref{KN3.19}. }
	\label{KN2}
\end{figure}
  In Fig.\ref{KN2} we plot the evolution of $H_i$ for the solution  \eqref{KN3.17}--\eqref{KN3.19} with \eqref{KN3.16}. It is interesting to study the evolution of the volume of the trefoil knot universe. For our case it is given by
  \begin{equation}\label{KN3.20}
V=[2+\cos (3\tau)]^2\cos (2\tau)\sin (2\tau)\sin (3\tau).
\end{equation}
\begin{figure}[h]
	\centering
		\includegraphics{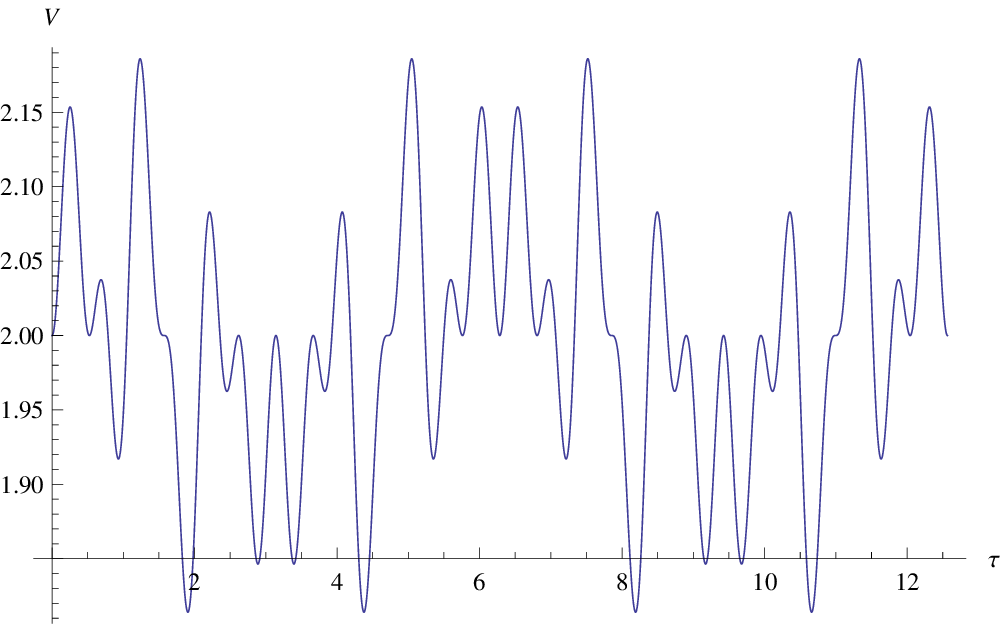}
	\caption{The  evolution of the volume of the trefoil knot universe with respect to the cosmic time $\tau$ for  Eq.\eqref{KN3.20}.}
	\label{KN3}
\end{figure}
 In Fig. \ref{KN3} we plot the  evolution of the volume of the trefoil knot universe with respect to the cosmic time $\tau$ for Eq. \eqref{KN3.20} with \eqref{KN3.16}. To  get $V\geq 0$, we must consider   $A_0, B_0, C_0>0$, if exactly  e.g. as $A_0>3, B_0>3, C_0>1$. But below for simplicity we take the case (3.16).  The other interesting quantity is the scalar curvature. For the trefoil knot solution (3.13)-(3.15), it has the form
 \begin{eqnarray} \label{R1}
R&=&(6(12\sin^2(3\tau)\sin^2(2\tau)-28\sin(3\tau)\cos(2\tau)\sin(2\tau)+3\sin^3(3\tau)\cos(2\tau)\sin(2\tau)-\notag\\& &
-12\sin^2(3\tau)\cos^2(2\tau)-8\cos(3\tau)\sin^2(2\tau)-8\cos^2(3\tau)\sin^2(2\tau)-2\cos^3(3\tau)\sin^2(2\tau)+\notag\\& &
+8\cos(3\tau)\cos^2(2\tau)+8\cos^2(3\tau)\cos^2(2\tau)+2\cos^3(3\tau)\cos^2(2\tau)-\notag\\& &
-52\sin(3\tau)\cos(2\tau)\cos(3\tau)\sin(2\tau)-19\cos(2\tau)\sin(2\tau)\sin(3\tau)\cos^2(3\tau)+\notag\\& &
+6\sin^2(2\tau)\sin^2(3\tau)\cos(3\tau)-\notag\\& &
-6\cos^2(2\tau)\sin^2(3\tau)\cos(3\tau)))/(\sin(2\tau)\cos(2\tau)(2+\cos(3\tau))^2\sin(3\tau)).
\end{eqnarray}

\begin{figure}[h]
	\centering
		\includegraphics[width=0.5 \textwidth]{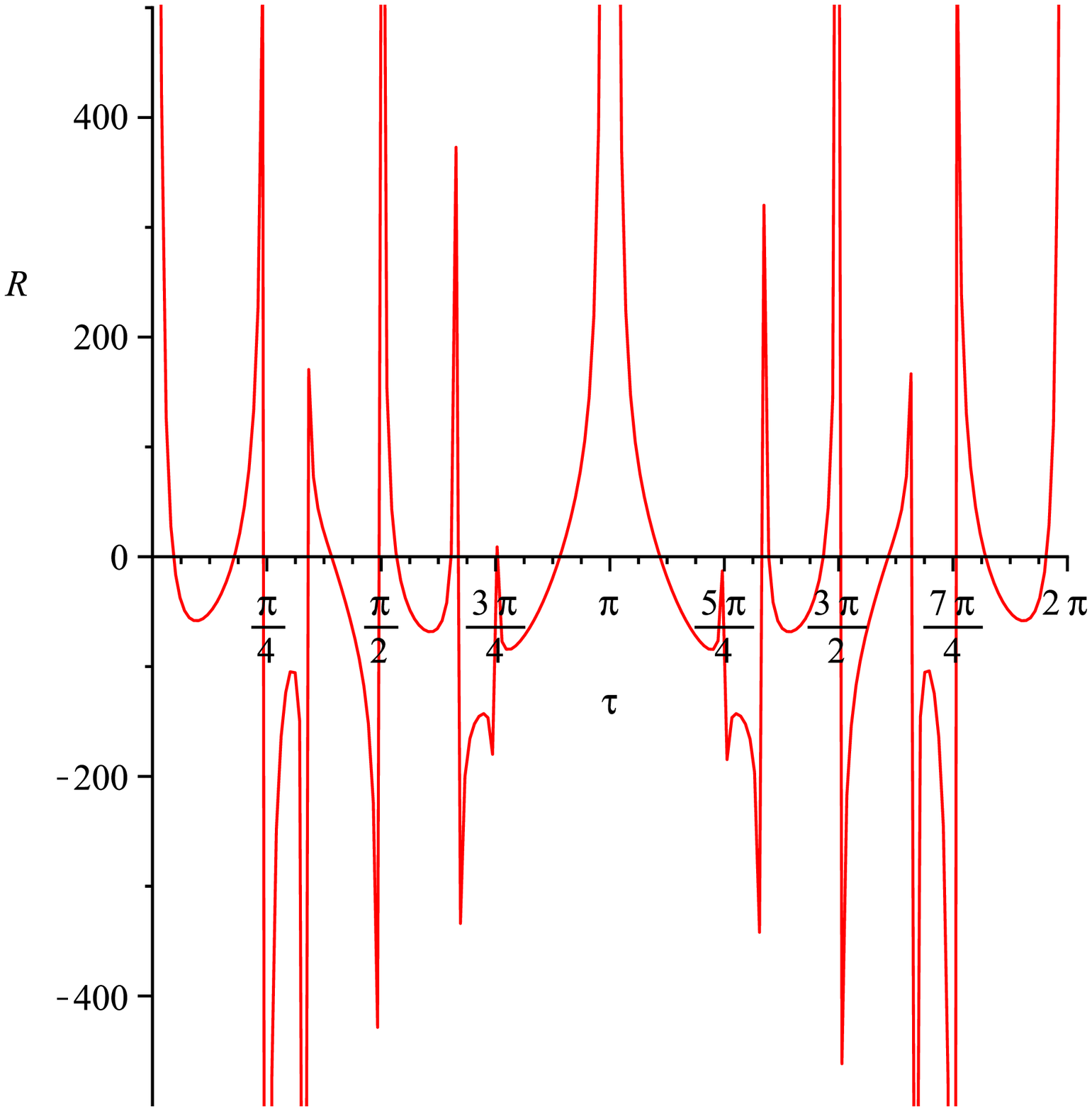}
	\caption{The evolution of the $R$ with respect of the cosmic time $\tau$ for  Eq.\eqref{R1}}
	\label{fig:R3.1.1}
\end{figure}
In Fig.\ref{fig:R3.1.1} we plot the evolution of the $R$ with respect of the cosmic time $\tau$.

So we have shown that the universe can live in the trefoil knot orbit according to the solution (3.13)-(3.15).  It is interesting to note that this trefoil knot solution admits infinite number accelerated and decelerated expansion phases of the universe. To show this, as an example let us consider the solution for $C$ from (3.13)-(3.15) that is $C=C_0+\sin (3\tau)$. In this case we have $\ddot{C}=-9\sin (3\tau)$ so that $\ddot{C}>0$ (accelerating phase)  as $\tau\in(\frac{\pi}{3}+\frac{2n\pi}{3},\frac{2\pi}{3}+\frac{2n\pi}{3})$ and $\ddot{C}<0$ (decelerating phase) as $\tau\in(\frac{2n\pi}{3}, \frac{\pi}{3}+\frac{2n\pi}{3})$ with the transion points $\dot{C}=3\cos(3\tau_i)=0$ as $\tau_i=(0.5 \pi +n\pi)/3$, where $n$ is integer that is $n=0, \pm 1, \pm 2, \pm 3, ...$.

 \subsection{Example 2.} 
\begin{figure}[h]
	\centering
		\includegraphics{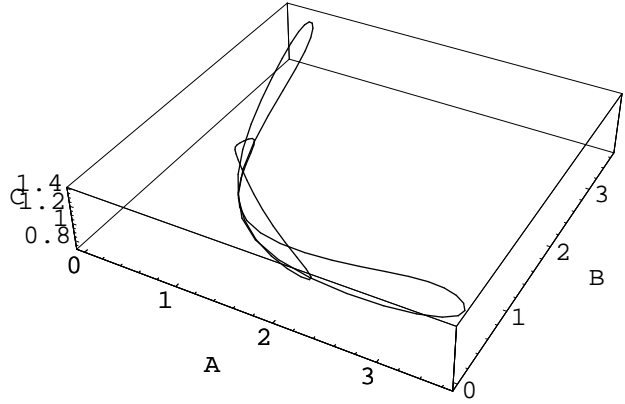}
	\caption{The evolution of  $A,B,C$ accordingly to \eqref{KN3.37}--\eqref{KN3.39}, [$t \in 0..2\pi$].}
	\label{fig:4}
\end{figure}

\begin{figure}[h]
	\centering
		\includegraphics{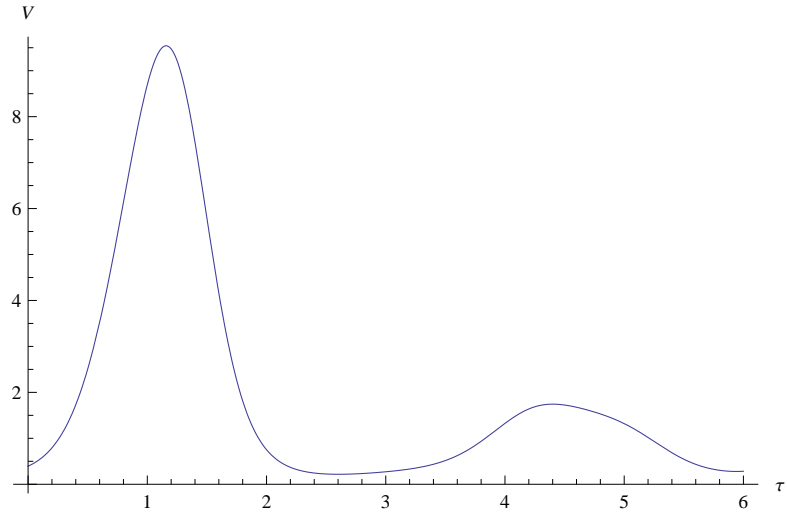}
	\caption{The  evolution of the volume for \eqref{KN3.40} with $A_0=B_0=C_0=V_0=1$.}
	\label{fig:5}
\end{figure}

Now we consider the following parametric  EoS 
\begin{eqnarray}\label{KN3.22}
p_1&=&-\frac{D_1}{E_1},\\
p_2&=&-\frac{D_2}{E_2},\\
p_3&=&-\frac{D_3}{E_3},\\
\rho&=&\frac{D_0}{E_0},\label{KN3.25}
\end{eqnarray}
where
\begin{eqnarray}
D_1&=&-\sin^2(2\tau)\cos^2(3\tau)+(-2\cos(2\tau)-4\sin^2(2\tau)
-3-\sin(3\tau)\sin(2\tau))\cos(3\tau)+\notag\\&
&
+\sin(3\tau)\sin(2\tau)-4\cos(2\tau)-\sin^2(3\tau)-4\sin^2(2\tau),\\
E_1&=&1,\\
D_2&=&-(2+\cos^2(3\tau))\cos^2(2\tau)-\sin(3\tau)(-1+\cos(3\tau))\cos(2\tau)+\notag\\&
&
+(-3+2\sin(2\tau))\cos(3\tau)+4\sin(2\tau)-\sin^2(3\tau),\\
E_2&=&1,\\
D_3&=&-(2+\cos^2(3\tau))\cos^2(2\tau)+(-\sin(2\tau)\cos^2(3\tau)+(-4\sin(2\tau)-2)\cos(3\tau)+3\sin(3\tau)-\notag\\&
&
-4-
4\sin(2\tau))\cos(2\tau)+2\sin(2\tau)\cos(3\tau)+(4+3\sin(3\tau))\sin(2\tau)-\sin^2(3\tau),\\
E_3&=&1,\\
D_0&=&(2+\cos(3\tau))(((2+\cos(3\tau))
\sin(2\tau)+\sin(3\tau))\cos(2\tau)+\sin(3\tau)\sin(2\tau)),\\
E_0&=&1.
\end{eqnarray}
Substituting these expressions for the pressures and  density of energy into the system \eqref{KN3.22}--\eqref{KN3.25}, we obtain the following its solution 
\begin{eqnarray}
H_1&=&[2+\cos (3\tau)]\cos (2\tau)=2\cos (2\tau)+0.5[\cos (5\tau)+\cos (\tau)], \\
H_2&=&[2+\cos (3\tau)]\sin (2\tau)=2\sin (2\tau)+0.5[\sin(5\tau)-\sin (\tau)], \\
H_3&=&\sin (3\tau).
    \end{eqnarray}
We see that this solution again describes the trefoil knot but for the "coordinates" $H_i$. Note that the scale factors we can   recovered from \eqref{KN2.17}. We get
   \begin{eqnarray}\label{KN3.37}
A&=&A_0e^{\sin(2\tau)+0.1\sin(5\tau)+0.5\sin(\tau)}, \\
B&=&B_0e^{-[\cos(2\tau)+0.1\cos(5\tau)-0.5\cos(\tau)]}, \\
C&=&C_0e^{-\frac{1}{3}\cos(3\tau)},\label{KN3.39}
\end{eqnarray}
where $A_0, B_0, C_0$ are some real constants. In Fig.\ref{fig:4} we plot the evolution of  $A,B,C$ accordingly to \eqref{KN3.37}--\eqref{KN3.39} and for the initial conditions $A(0)=1, B(0)=e^{-0.6}, C(0)=e^{-1/3}$, where  we assume that $A_0=B_0=C_0=1$.
For this example,  the volume of the universe is given by
  \begin{equation}\label{KN3.40}
V=V_0e^{\{\sin(2\tau)+0.1\sin(5\tau)+0.5\sin(\tau)-[\cos(2\tau)+0.1\cos(5\tau)-0.5\cos(\tau)]-\frac{1}{3}\cos(3\tau)\}}.
\end{equation}
The  evolution of the volume for \eqref{KN3.40} is  presented in Fig. \ref{fig:5} for $A_0=B_0=C_0=V_0=1$ and for the intial condition $V(0)=e^{-14/15}$.

 The scalar curvature has the form
 \begin{eqnarray}\label{R2}
R&=&(2\cos^2(2\tau)+2\sin^2(2\tau)+2\cos(2\tau)\sin(2\tau))\cos^2(3\tau)+(8\cos^2(2\tau)+(2\sin(3\tau)+4+\notag\\& &
+8\sin(2\tau))\cos(2\tau)+6+8\sin^2(2\tau)+(-4+2\sin(3\tau))\sin(2\tau))\cos(3\tau)+8\cos^2(2\tau)+\notag\\& &
+(-2\sin(3\tau)+8+8\sin(2\tau))\cos(2\tau)+8\sin^2(2\tau)+\notag\\& &
+(-8-2\sin(3\tau))\sin(2\tau)+2\sin^2(3\tau).
\end{eqnarray}
\begin{figure}[h]
	\centering
		\includegraphics[width=0.5 \textwidth]{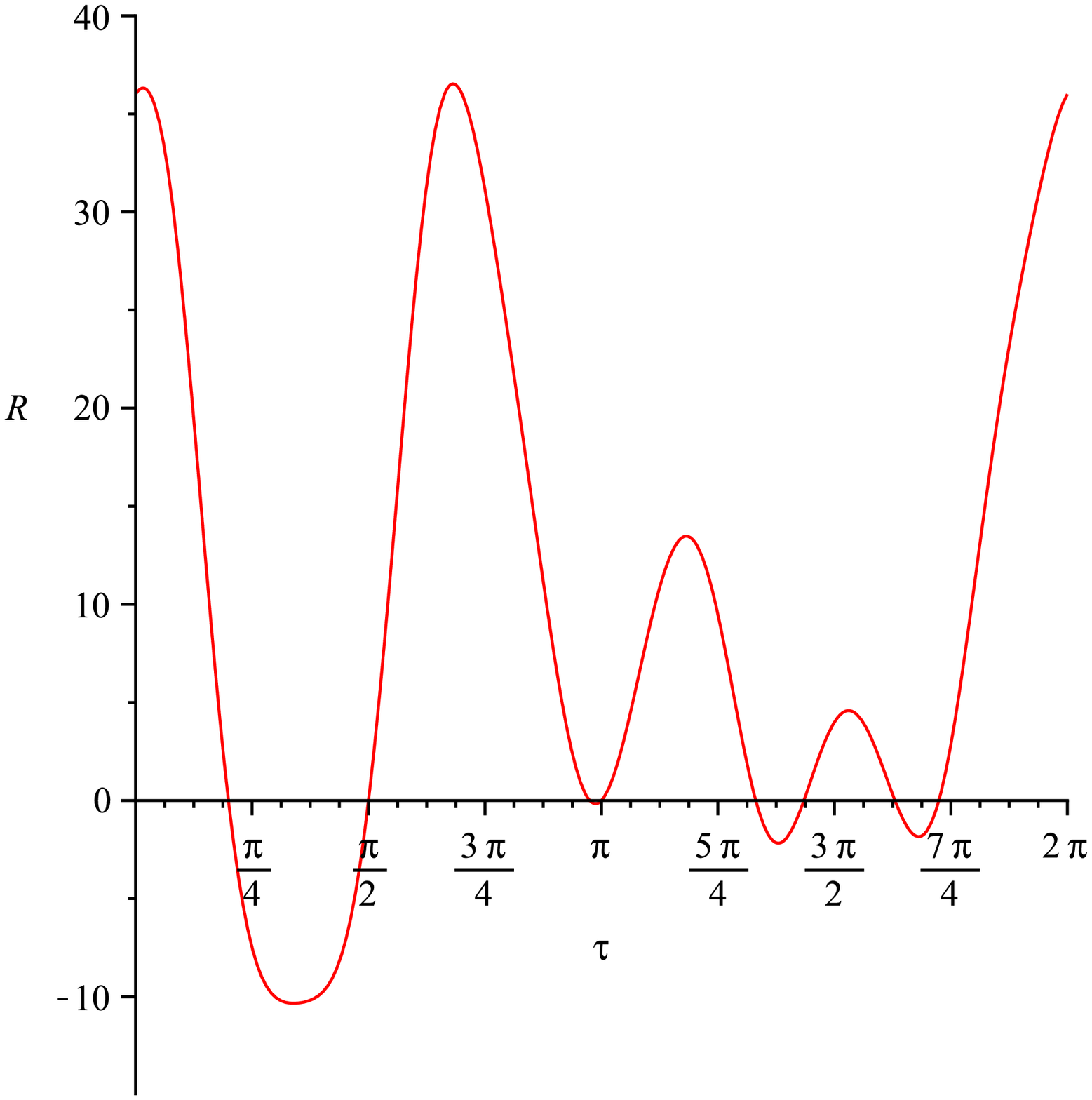}
	\caption{The evolution of the $R$ with respect of the cosmic time $\tau$ for  Eq.\eqref{R2}}
	\label{fig:R3.2.2}
\end{figure}
In Fig.\ref{fig:R3.2.2} we plot the evolution of the $R$ with respect of the cosmic time $\tau$. Finally we conclude  that the Einstein equations for the Bianchi I type metric admit the trefoil knot solution of the form (3.34)-(3.36) or (3.37)-(3.39). These solutions describe the accelerated and decelerated phases of the expansion of the universe.

 \subsection{Example 3.} 
\begin{figure}[h]
	\centering
		\includegraphics{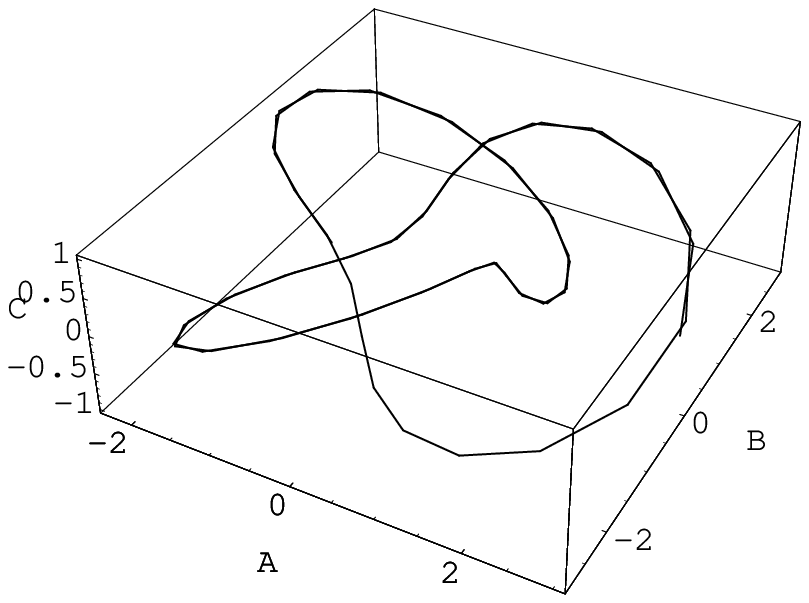}
		\caption{The knotted closed curve  corresponding to the solution \eqref{KN3.42}--\eqref{KN3.44} with \eqref{KN3.16}, $t \in [0,4\pi$], $k=1/3$.}
	\label{fig:6}
\end{figure}

\begin{figure}[h]
	\centering
		\includegraphics[width=0.5 \textwidth]{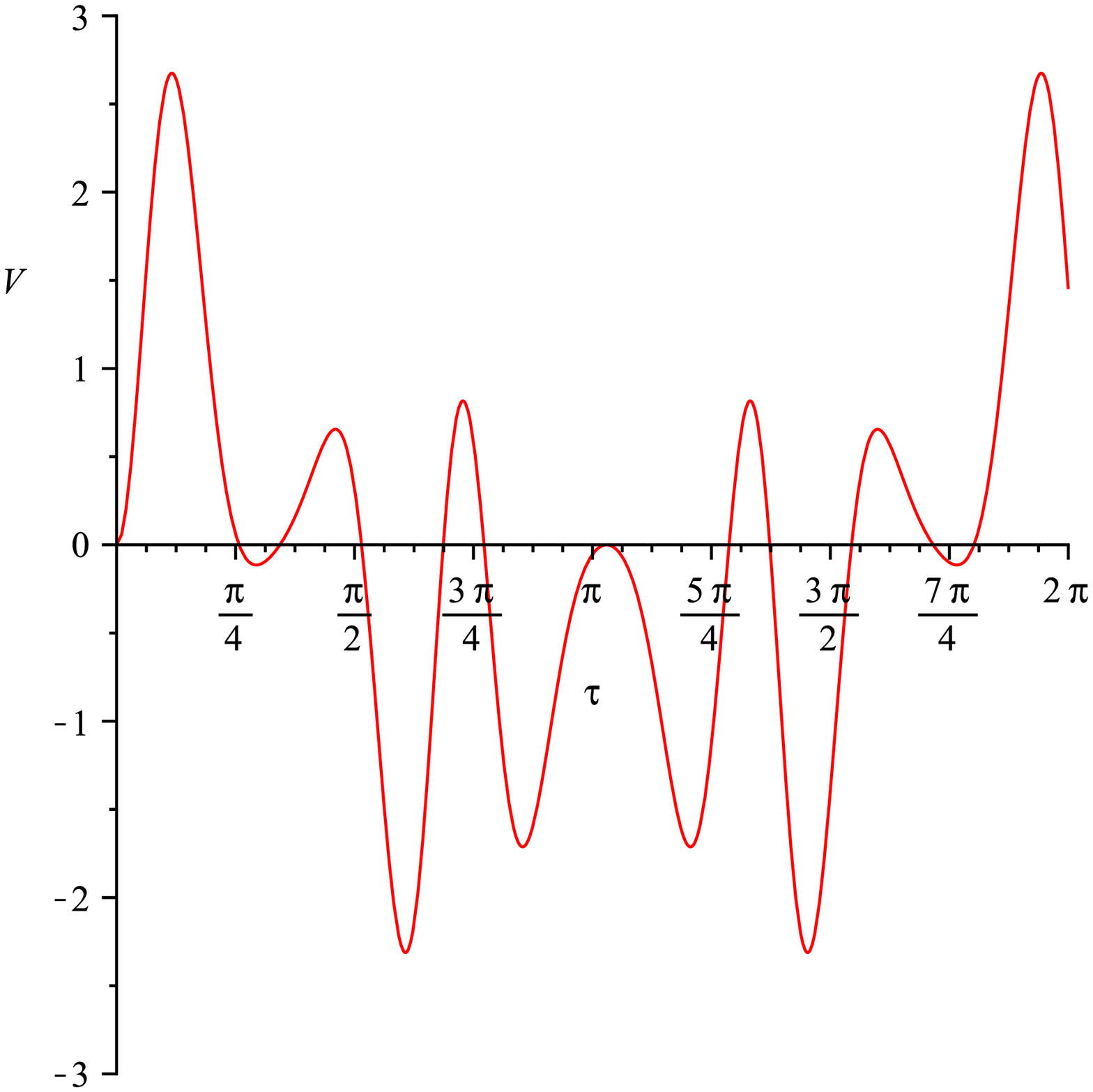}
	\caption{The  evolution of the volume of the trefoil knot universe with respect to the cosmic time $\tau$ for  Eq.\eqref{V3}}
	\label{fig:V3.3.3}
\end{figure}

 Now we  present  a new kind of  the trefoil knot universes. Let the system \eqref{KN2.21}--\eqref{KN2.24}  has the solution 
\begin{eqnarray}\label{KN3.42}
A&=&A_0+[2+\mbox{cn}(3\tau)]\mbox{cn}(2\tau), \\
B&=&B_0+[2+\mbox{cn}(3\tau)]\mbox{sn} (2\tau), \\
C&=&C_0+\mbox{sn}(3\tau),\label{KN3.44}
\end{eqnarray}
where $\mbox{cn}(t)\equiv\mbox{cn}(t,k)$ and $\mbox{sn}(t)\equiv\mbox{sn}(t,k)$ are  the Jacobian  elliptic functions which are  doubly periodic functions, $k$ is the elliptic modulus. Fig. \ref{fig:6} shows the knotted closed curve  corresponding to the solution \eqref{KN3.42}--\eqref{KN3.44} with \eqref{KN3.16}. Substituting the formulas \eqref{KN3.42}--\eqref{KN3.44} into the system \eqref{KN2.21}--\eqref{KN2.24} we get the corresponding expressions for $\rho$ and $p_i$ that gives us the parametric EoS. This parametric EoS reads as
\begin{eqnarray}
p_1&=&-\frac{D_1}{E_1},\\
p_2&=&-\frac{D_2}{E_2},\\
p_3&=&-\frac{D_3}{E_3},\\
\rho&=&\frac{D_0}{E_0},
\end{eqnarray}
where
\begin{eqnarray}
D_1&=&9k^2cn(3\tau, k)sn^3(3\tau, k)sn(2\tau, k)-12\frac{\partial}{\partial\tau}am(3\tau, k)sn^2(3\tau, k)cn(2\tau, k)\frac{\partial}{\partial\tau}am(2\tau, k)-\notag\\&
&-4sn(2\tau, k)((9/4)cn^3(3\tau, k)k^2+(9/2)cn^2(3\tau, k)k^2+\notag\\&
&+((45/4)\frac{\partial}{\partial\tau}am^2(3\tau, k)+cn^2(2\tau, k)k^2+\frac{\partial}{\partial\tau}am^2(2\tau, k))cn(3\tau, k)+2\frac{\partial}{\partial\tau}am^2(2\tau, k)+\notag\\&
&+(9/2)\frac{\partial}{\partial\tau}am^2(3\tau, k)+2cn^2(2\tau, k)k^2)sn(3\tau, k)+18cn(3\tau, k)\frac{\partial}{\partial\tau}am(3\tau, k)\notag\\&
&(2+cn(3\tau, k))cn(2\tau, k)\frac{\partial}{\partial\tau}am(2\tau, k),\\
E_1&=&(2+cn(3\tau, k))sn(2\tau, k)sn(3\tau, k),\\
D_2&=&9k^2cn(3\tau, k)sn^3(3\tau, k)cn(2\tau, k)+12\frac{\partial}{\partial\tau}am(3\tau, k)sn^2(3\tau, k)\frac{\partial}{\partial\tau}am(2\tau, k)sn(2\tau, k)-\notag\\&
&-9cn(2\tau, k)cn^3(3\tau, k)k^2+2cn^2(3\tau, k)k^2+((4/9)\frac{\partial}{\partial\tau}am^2(2\tau, k)+5\frac{\partial}{\partial\tau}am^2(3\tau, k)-\notag\\&
&-(4/9)k^2sn^2(2\tau, k))cn(3\tau, k)+(8/9)\frac{\partial}{\partial\tau}am^2(2\tau, k)+2\frac{\partial}{\partial\tau}am^2(3\tau, k)-\notag\\&
&-(8/9)k^2sn^2(2\tau, k))sn(3\tau, k)-\notag\\&
&-18cn(3\tau, k)\frac{\partial}{\partial\tau}am(3\tau, k)(2+cn(3\tau, k))\frac{\partial}{\partial\tau}am(2\tau, k)sn(2\tau, k),\\
E_2&=&sn(3\tau, k)(2+cn(3\tau, k))cn(2\tau, k),\\
D_3&=&-4k^2sn(2\tau, k)(2+cn(3\tau, k))^2cn^3(2\tau, k)-30\frac{\partial}{\partial\tau}am(2\tau, k)sn(3\tau, k)\frac{\partial}{\partial\tau}am(3\tau, k)\notag\\&
&(2+cn(3\tau, k))cn^2(2\tau, k)+4sn(2\tau, k)(k^2(2+cn^2(3\tau, k))sn^2(2\tau, k)+\notag\\&
&+(-(9/2)\frac{\partial}{\partial\tau}am^2(3\tau, k)+(9/2)k^2sn^2(3\tau, k)-5\frac{\partial}{\partial\tau}am^2(2\tau, k))cn^2(3\tau, k)+\notag\\&
&+(-20\frac{\partial}{\partial\tau}am^2(2\tau, k)+9k^2sn^2(3\tau, k)-9\frac{\partial}{\partial\tau}am^2(3\tau, k))cn(3\tau, k)+\notag\\&
&+(27/4)sn^2(3\tau, k)\frac{\partial}{\partial\tau}am(3\tau, k)^2-20\frac{\partial}{\partial\tau}am^2(2\tau, k))cn(2\tau, k)+\notag\\&
&+30\frac{\partial}{\partial\tau}am(3\tau, k)sn(3\tau, k)\frac{\partial}{\partial\tau}am(2\tau, k)sn^2(2\tau, k)(2+cn(3\tau, k)),\\
E_3&=&(2+cn(3\tau, k))^2cn(2\tau, k)sn(2\tau, k),\\
D_0&=&-4sn(3\tau, k)sn(2\tau, k)cn(2\tau, k)(2+cn(3\tau, k))^2\frac{\partial}{\partial\tau}am^2(2\tau, k)+6\frac{\partial}{\partial\tau}am(\tau, k)(2+cn(3\tau, k))\notag\\&
&(cn^2(3\tau, k)+2cn(3\tau, k)-sn^2(3\tau, k))(cn(2\tau, k)-sn(2\tau, k))(cn(2\tau, k)+sn(2\tau, k))\notag\\&
&\frac{\partial}{\partial\tau}am(2\tau, k)-18sn(3\tau, k)sn(2\tau, k)(-(1/2)sn^2(3\tau, k)+cn^2(3\tau, k)+\notag\\&
&+2cn(3\tau, k))cn(2\tau, k)\frac{\partial}{\partial\tau}am^2(3\tau, k),\\
E_0&=&(2+cn^2(3\tau, k))cn(2\tau, k)sn(2\tau, k)sn(3\tau, k).
\end{eqnarray}
The  volume of the universe for the solution \eqref{KN3.42}--\eqref{KN3.44} with \eqref{KN3.16} looks like
  \begin{equation}\label{V3}
V=[2+\mbox{cn}(3\tau)]^2\mbox{cn}(2\tau)\mbox{sn}(2\tau)\mbox{sn}(3\tau).
\end{equation}
The  evolution of the volume for \eqref{V3} is  presented in Fig.\ref{fig:V3.3.3}
 The scalar curvature has the form
 \begin{eqnarray} \label{R3}
R&=&(-8\mbox{sn}(2\tau, k)\mbox{sn}(3\tau, k)k^2(2+\mbox{cn}(3\tau, k))^2\mbox{cn}^3(2\tau, k)+\notag\\& &
+12\mbox{dn}(2\tau, k)\mbox{dn}(3\tau, k)(2+\mbox{cn}(3\tau, k))(\mbox{cn}^2(3\tau, k)+\notag\\& &
+2\mbox{cn}(3\tau, k)-3\mbox{sn}^2(3\tau, k))\mbox{cn}^2(2\tau, k)-18\mbox{sn}(3\tau, k)(-(4/9)k^2(2+\mbox{cn}(3\tau, k))^2\mbox{sn}^2(2\tau, k)+\notag\\& &
+(-4k^2\mbox{cn}(3\tau, k)-2\mbox{cn}^2(3\tau, k)k^2-\mbox{dn}^2(3\tau, k))\mbox{sn}^2(3\tau, k)+\notag\\& &
+(2+\mbox{cn}(3\tau, k))(\mbox{cn}^3(3\tau, k)k^2+2\mbox{cn}^2(3\tau, k)k^2+\notag\\& &
+(5\mbox{dn}^2(3\tau, k)+(4/3)\mbox{dn}^2(2\tau, k))\mbox{cn}(3\tau, k)+\notag\\& &+(8/3)\mbox{dn}^2(2\tau, k)+2\mbox{dn}^2(3\tau, k)))\mbox{sn}(2\tau, k)\mbox{cn}(2\tau, k)-\notag\\& &
-12\mbox{dn}(2\tau, k)\mbox{sn}^2(2\tau, k)\mbox{dn}(3\tau, k)(2+\mbox{cn}(3\tau, k))(\mbox{cn}^2(3\tau, k)+\notag\\& &
+2\mbox{cn}(3\tau, k)-3\mbox{sn}^2(3\tau, k)))/(\mbox{cn}(2\tau, k)\mbox{sn}(2\tau, k)(2+\mbox{cn}(3\tau, k))^2\mbox{sn}(3\tau, k)).
\end{eqnarray}
\begin{figure}[h]
	\centering
		\includegraphics[width=0.7 \textwidth]{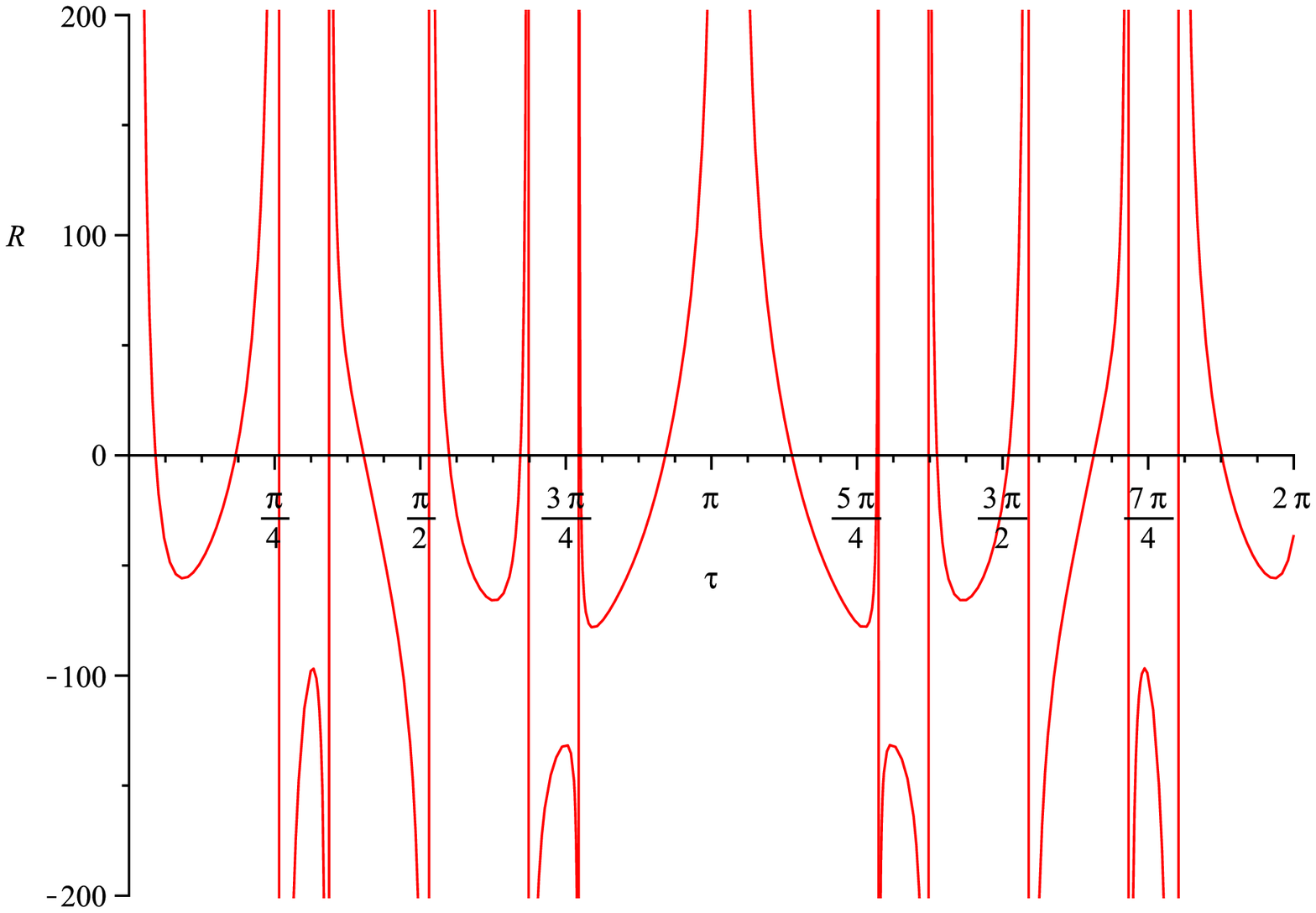}
	\caption{The evolution of the $R$ with respect of the cosmic time $\tau$  for  Eq.\eqref{R3}}
	\label{fig:R3.3.3}
\end{figure}
In Fig.\ref{fig:R3.3.3} we plot the evolution of the $R$ with respect of the cosmic time $\tau$.
     \subsection{Example 4.} Our fourth example is given by 
    \begin{eqnarray}
H_1&=&[2+\mbox{cn}(3\tau)]\mbox{cn}(2\tau), \\
H_2&=&[2+\mbox{cn}(3\tau)]\mbox{sn} (2\tau), \\
H_3&=&\mbox{sn}(3\tau)
    \end{eqnarray}which again the  knotted closed curve in Fig. \ref{fig:6} but for the "coordinates" $H_i$.  Note that the corresponding parametric EoS looks like
 \begin{eqnarray}\label{KN3.62}
p_1&=&-\frac{D_1}{E_1},\\
p_2&=&-\frac{D_2}{E_2},\\
p_3&=&-\frac{D_3}{E_3},\\
\rho&=&\frac{D_0}{E_0},\label{KN3.65}
 \end{eqnarray}
where
\begin{eqnarray}
D_1&=&-(2+cn(3\tau, k))^2sn^2(2\tau, k)-sn(3\tau, k)(-3\frac{\partial}{\partial\tau}am(3\tau, k)+2+cn(3\tau, k))sn(2\tau, k)+\notag\\&
&+(-2cn(2\tau, k)\frac{\partial}{\partial\tau}am(2\tau, k)-3\frac{\partial}{\partial\tau}am(3\tau, k))cn(3\tau, k)-4cn(2\tau, k)\frac{\partial}{\partial\tau}am(2\tau, k)-\notag\\&
&-sn^2(3\tau, k),\\
E_1&=&1,\\
D_2&=&-(2+cn(3\tau, k))^2cn^2(2\tau, k)-sn(3\tau, k)(-3\frac{\partial}{\partial\tau}am(3\tau, k)+2+cn(3\tau, k))cn(2\tau, k)+\notag\\&
&+(-3\frac{\partial}{\partial\tau}am(3\tau, k)+2\frac{\partial}{\partial\tau}am(2\tau, k)sn(2\tau, k))cn(3\tau, k)+4\frac{\partial}{\partial\tau}am(2\tau, k)sn(2\tau, k)-\notag\\&
&-sn^2(3\tau, k),\\
E_2&=&1,\\
D_3&=&-(2+cn(3\tau, k))^2cn^2(2\tau, k)+(-sn(2\tau, k)cn^2(3\tau, k)+(-4sn(2\tau, k)-2\frac{\partial}{\partial\tau}am(2\tau, k))\notag\\&
&cn(3\tau, k)+3sn(3\tau, k)\frac{\partial}{\partial\tau}am(3\tau, k)-4\frac{\partial}{\partial\tau}am(2\tau, k)-4sn(2\tau, k))cn(2\tau, k)+\notag\\&
&+2\frac{\partial}{\partial\tau}am(2\tau, k)sn(2\tau, k)cn(3\tau, k)+(4\frac{\partial}{\partial\tau}am(2\tau, k)+3sn(3\tau, k)\frac{\partial}{\partial\tau}am(3\tau, k))sn(2\tau, k)-\notag\\&
&-sn^2(3\tau, k),\\
E_3&=&1,\\
D_0&=&(((2+cn(3\tau, k))sn(2\tau, k)+sn(3\tau, k))cn(2\tau, k)+sn(2\tau, k)sn(3\tau, k))\notag\\&
&(2+cn(3\tau, k)),\\
E_0&=&1.
\end{eqnarray}
\begin{figure}[h]
	\centering
		\includegraphics{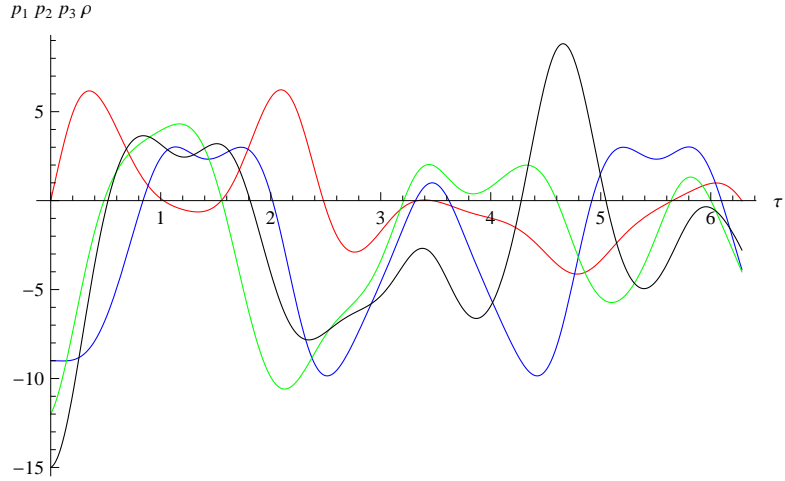}
\caption{The evolution of $p_i, \rho$ for \eqref{KN3.62}--\eqref{KN3.65}, $t \in [0,2\pi$], $k=1/3$, $\rho$(red), $p_1$(blue), $p_2$(green), $p_3$(black).}
\label{fig:7}	
\end{figure}
In Fig. \ref{fig:7} we plot the evolution of $p_i, \rho$ for \eqref{KN3.62}--\eqref{KN3.65}. 
 The scalar curvature has the form
 \begin{eqnarray}\label{R4}
R&=&2(2+\mbox{cn}(3\tau, k))^2\mbox{cn}^2(2\tau, k)+(2\mbox{sn}(2\tau, k)\mbox{cn}^2(3\tau, k)+(8\mbox{sn}(2\tau, k)+4\mbox{dn}(2\tau, k)+\notag\\& &
+2\mbox{sn}(3\tau, k))\mbox{cn}(3\tau, k)+8\mbox{sn}(2\tau, k)+(4-6\mbox{dn}(3\tau, k))\mbox{sn}(3\tau, k)+\notag\\& &
+8\mbox{dn}(2\tau, k))\mbox{cn}(2\tau, k)+2\mbox{sn}^2(2\tau, k)\mbox{cn}^2(3\tau, k)+(8\mbox{sn}^2(2\tau, k)+\notag\\& &
+(-4\mbox{dn}(2\tau, k)+2\mbox{sn}(3\tau, k))\mbox{sn}(2\tau, k)+6\mbox{dn}(3\tau, k))\mbox{cn}(3\tau, k)+\notag\\& &
+8\mbox{sn}^2(2\tau, k)+((4-6\mbox{dn}(3\tau, k))\mbox{sn}(3\tau, k)-8\mbox{dn}(2\tau, k))\mbox{sn}(2\tau, k)+2\mbox{sn}(3\tau, k)^2
\end{eqnarray}
\begin{figure}[h]
	\centering
		\includegraphics[width=0.5 \textwidth]{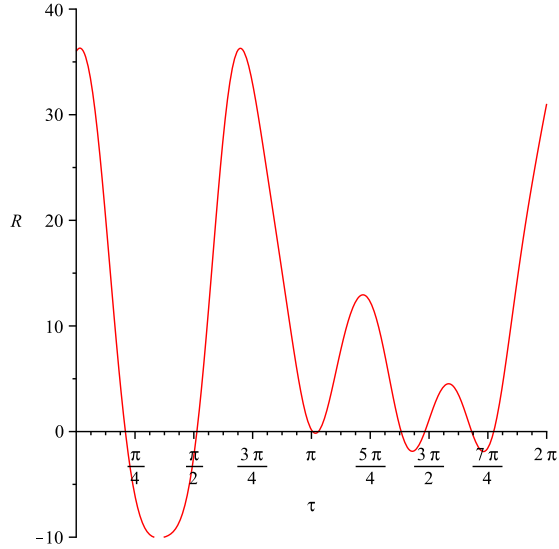}
	\caption{The evolution of the $R$ with respect of the cosmic time $\tau$ for  Eq.\eqref{R4}}
	\label{fig:R3.4.4}
\end{figure}
In Fig.\ref{fig:R3.4.4} we plot the evolution of the $R$ with respect of the cosmic time $\tau$.
   \section{The figure-eight knot universe}
  Our aim in this section is to  demonstrate some examples the  figure-eight knot universes for the Bianchi type I metric \eqref{KN2.6}. We give some particular figure-eight knot universe models.

   \subsection{Example 1.} Again, let us  we assume that our universe is filled by the fluid with the following parametric EoS
\begin{eqnarray}
\rho&=&\frac{D_8}{E_8},\\
p_1&=&-\frac{D_9}{E_9},\\
p_2&=&-\frac{D_{10}}{E_{10}},\\
p_3&=&-\frac{D_{11}}{E_{11}},
\end{eqnarray}
where
\begin{eqnarray}
D_8&=&(-2\sin(2\tau)\cos(3\tau)-(3(2+\cos(2\tau)))\sin(3\tau))(-2\sin(2\tau)\sin(3\tau)+\notag\\
& &(3(2+\cos(2\tau)))\cos(3\tau))\sin(4\tau)+12\cos(3\tau)((2+\cos(2\tau))\cos(3\tau)-\notag\\
& &(2/3)\sin(2\tau)\sin(3\tau))(2+\cos(2\tau))\cos(4\tau)-(12(2+\cos(2\tau)))\cos(4\tau)*\notag\\
& &((2+\cos(2\tau))\sin(3\tau)+(2/3)\sin(2\tau)\cos(3\tau))\sin(3\tau),\\
E_8&=&(2+\cos(2\tau))^2\cos(3\tau)\sin(3\tau)\sin(4\tau),\\
D_9&=&((72+36\cos(2\tau))\cos(4\tau)-12\sin(4\tau)\sin(2\tau))\cos(3\tau)-(29((24/29)*\notag\\
& &\cos(4\tau)\sin(2\tau)+(cos(2\tau)+50/29)\sin(4\tau)))\sin(3\tau),\\
E_9&=&\sin(4\tau)(2+\cos(2\tau))\sin(3\tau),\\
D_{10}&=&(-24\cos(4\tau)\sin(2\tau)+\sin(4\tau)(-29\cos(2\tau)-50))\cos(3\tau)-\notag\\
& &(36((2+\cos(2\tau))\cos(4\tau)-(1/3)\sin(4\tau)\sin(2\tau)))\sin(3\tau),\\
E_{10}&=&\sin(4\tau)(2+\cos(2\tau))\cos(3\tau),\\
D_{11}&=&-(30(2+\cos(2\tau)))\sin(2\tau)\cos(3\tau)^2+\sin(3\tau)(12\sin(2\tau)^2-196\cos(2\tau)-\notag\\
& &180-53\cos(2\tau)^2)\cos(3\tau)+30\sin(2\tau)\sin(3\tau)^2(2+\cos(2\tau)),\\
E_{11}&=&(2+\cos(2\tau))^2\cos(3\tau)\sin(3\tau).
\end{eqnarray}

\begin{figure}[h]
	\centering
		\includegraphics{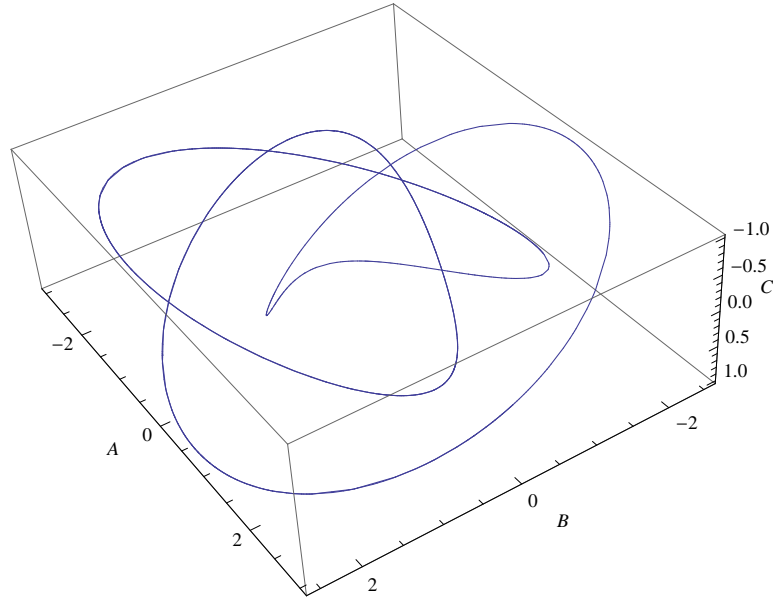}
	\caption{The figure-eight knot for \eqref{KN4.13}--\eqref{KN4.15} with \eqref{KN3.16}.}
	\label{fig:8}
\end{figure}

Substituting these expressions for the pressuries and the density of energy into the system \eqref{KN2.21}--\eqref{KN2.24}, we obtain the following its solution \cite{Knot-universe}
\begin{eqnarray}\label{KN4.13}
A&=&A_0+[2+\cos (2\tau)]\cos (3\tau), \\
B&=&B_0+[2+\cos (2\tau)]\sin (3\tau), \\
C&=&C_0+\sin (4\tau).\label{KN4.15}
    \end{eqnarray}
 This solution is nothing but the parametric equation of the figure-eight knot as we can see from Fig. \ref{fig:8}, where we assume that $A_0=B_0=C_0=0$ and the initial conditions have the form $A(0)=3, B(0)=0, C(0)=0$.  And for that reason  in \cite{Knot-universe} we called  such models as the figure-eight knot universes.
Note that the "coordinates" $A,B,C$ with (3.16) satisfy the equation
    \begin{equation}
4(h-2)^4-4(h-2)^2+z^2=0,
\end{equation}where $h=2+\cos (2\tau)$. 
  Let us  calculate the volume of the universe. For our case it is given by
  \begin{equation}
V=[2+\cos(2\tau)]^2\cos(3\tau)\sin(3\tau)\sin(4\tau),
\end{equation}
where we used \eqref{KN3.16}.
\begin{figure}[h]
	\centering
		\includegraphics{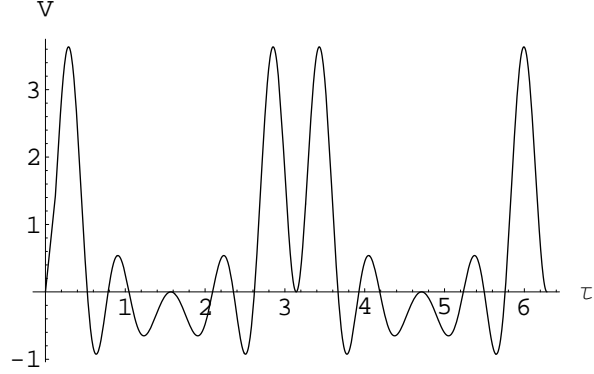}
		\caption{The evolution of the volume for the solution \eqref{KN4.13}--\eqref{KN4.15} with \eqref{KN3.16}, $t \in [0,2\pi$].}
	\label{fig:9}
\end{figure}
In Fig.9 we present   the evolution of the volume for the solution \eqref{KN4.13}--\eqref{KN4.15} with \eqref{KN3.16}.  
 The scalar curvature has the form
 \begin{eqnarray}\label{R5}
R&=&((24(\cos(4\tau)\cos(2\tau)-(3/2)\sin(4\tau)\sin(2\tau)+2\cos(4\tau)))(2+\cos(2\tau))\cos^2(3\tau)-\notag\\& &
-102\sin(3\tau)(\sin(4\tau)\cos^2(2\tau)+((188/51)\sin(4\tau)+(16/51)\cos(4\tau)\sin(2\tau))\cos(2\tau)+\notag\\& &
+(32/51)\cos(4\tau)\sin(2\tau)+(172/51)\sin(4\tau)-\notag\\& & -(4/51)\sin^2(2\tau)\sin(4\tau))\cos(3\tau)-24\sin^2(3\tau)(\cos(4\tau)\cos(2\tau)-\notag\\& &
-(3/2)\sin(4\tau)\sin(2\tau)+2\cos(4\tau))(2+\cos(2\tau)))/(\sin(3\tau)\cos(3\tau)*\notag\\& &
*(2+\cos(2\tau))^2\sin(4\tau)).
\end{eqnarray}
\begin{figure}[h]
	\centering
		\includegraphics[width=0.5 \textwidth]{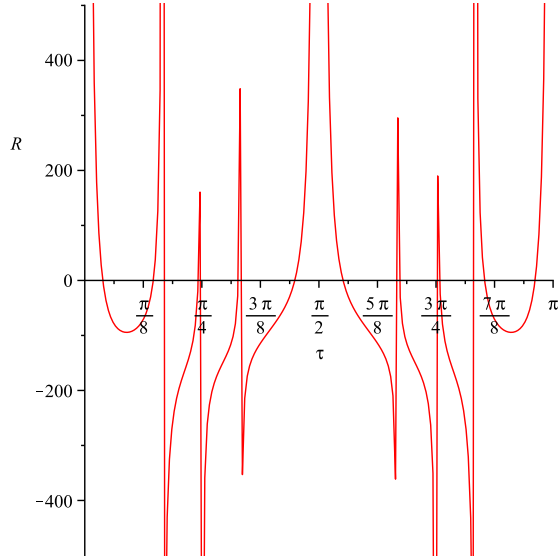}
	\caption{The evolution of the $R$ with respect of the cosmic time $\tau$ for  Eq.\eqref{R5}}
	\label{fig:R4.1.1}
\end{figure}
In Fig.\ref{fig:R4.1.1} we plot the evolution of the $R$ with respect of the cosmic time $\tau$.
So we found the figure-eight knot solution of the Einstein equations which again describe the accelerated and decelerated  expansion phases of the universe.
   \subsection{Example 2} Now we consider the system \eqref{KN2.25}--\eqref{KN2.28}. Let its  solution is given by
    \begin{eqnarray}
H_1&=&[2+\cos (2\tau)]\cos (3\tau)=2\cos (3\tau)+\cos (5\tau)+\cos (\tau), \\
H_2&=&[2+\cos (2\tau)]\sin (3\tau)=2\sin (3\tau)+\sin(\tau)+\sin (5\tau), \\
H_3&=&\sin (4\tau).
    \end{eqnarray}
Then  the coorresponding scale factors read as
 \begin{eqnarray}
A&=&A_0e^{\frac{2}{3}\sin(3\tau)+0.2\sin (5\tau)+\sin (\tau)}, \\
B&=&B_0e^{-[\frac{2}{3}\cos(3\tau)+0.2cos(5\tau)+\cos(\tau)]}, \\
C&=&C_0e^{-0.25\cos (4\tau)}.
    \end{eqnarray}
    For this solution the parametric EoS looks like
  \begin{eqnarray}\label{KN4.25}
\rho&=&\frac{D_0}{E_0},\\
p_1&=&-\frac{D_1}{E_1},\\
p_2&=&-\frac{D_2}{E_2},\\
p_3&=&-\frac{D_3}{E_3},\label{KN4.28}
\end{eqnarray}
where
\begin{eqnarray}
D_0&=&(((2+\cos(2\tau))\sin(3\tau)+\sin(4\tau))\cos(3\tau)+\sin(3\tau)\sin(4\tau))(2+\cos(2\tau)),\\
E_0&=&1,\\
D_1&=&-(2+\cos(2\tau))^2\sin^2(3\tau)+(2\sin(2\tau)-2\sin(4\tau)-\sin(4\tau)\cos(2\tau))\sin(3\tau)-\notag\\&
&-6\cos(3\tau)-3\cos(3\tau)\cos(2\tau)-4\cos(4\tau)-\sin^2(4\tau),\\
E_1&=&1,\\
D_2&=&-(2+\cos(2\tau))^2\cos^2(3\tau)+(2\sin(2\tau)-2\sin(4\tau)-\sin(4\tau)\cos(2\tau))\cos(3\tau)-\notag\\&
&-4\cos(4\tau)+6\sin(3\tau)+3\sin(3\tau)\cos(2\tau)-\sin^2(4\tau),\\
E_2&=&1,\\
D_3&=&-3\sin(\tau)-64\sin(\tau)\cos^9(\tau)+36\sin(\tau)\cos^5(\tau)+40\sin(\tau)\cos^4(\tau)+4\sin(\tau)\cos^3(\tau)-\notag\\&
&-6\sin(\tau)\cos^2(\tau)-3\sin(\tau)\cos(\tau)-25\cos^2(\tau)+5\cos(\tau)-40\cos^5(\tau)-64\cos^10(\tau)+\notag\\&
&+96\cos^8(\tau)-84\cos^6(\tau)+68\cos^4(\tau)+26\cos^3(\tau),\\
E_3&=&1.
\end{eqnarray}
\begin{figure}[h]
	\centering
		\includegraphics{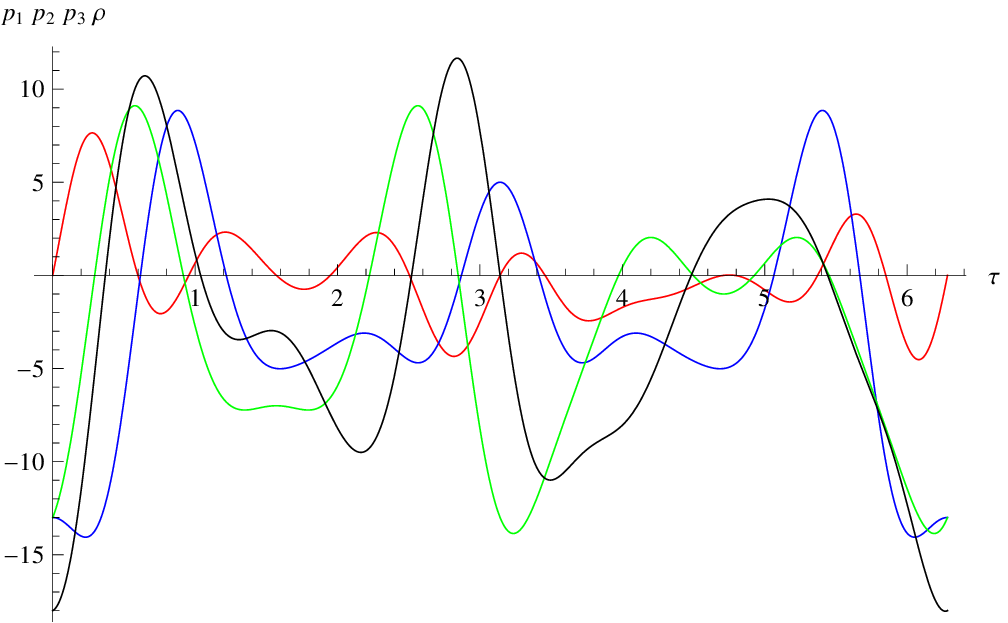}
		\caption{ The plot of  the EoS \eqref{KN4.25}--\eqref{KN4.28}, $t \in [0, 2\pi$], $\rho$(red), $p_1$(blue), $p_2$(green), $p_3$(black).}
	\label{fig:10}
\end{figure}
In  Fig. \ref{fig:10} we  plot  the EoS \eqref{KN4.25}--\eqref{KN4.28}.
For this example, the evolution of the volume of the universe is given by
  \begin{equation}\label{KN4.37}
V=V_0e^{\frac{2}{3}\sin(3\tau)+0.2\sin (5\tau)+\sin (\tau)-\frac{2}{3}\cos(3\tau)-0.2cos(5\tau)-\cos(\tau)-0.25\cos (4\tau)}.
\end{equation}
The  evolution of the volume is presented in Fig.\ref{fig:11} for $A_0=B_0=C_0=V_0=1$ and for the intial condition $V(0)=e^{-127/60}$.
\begin{figure}[h]
	\centering
		\includegraphics{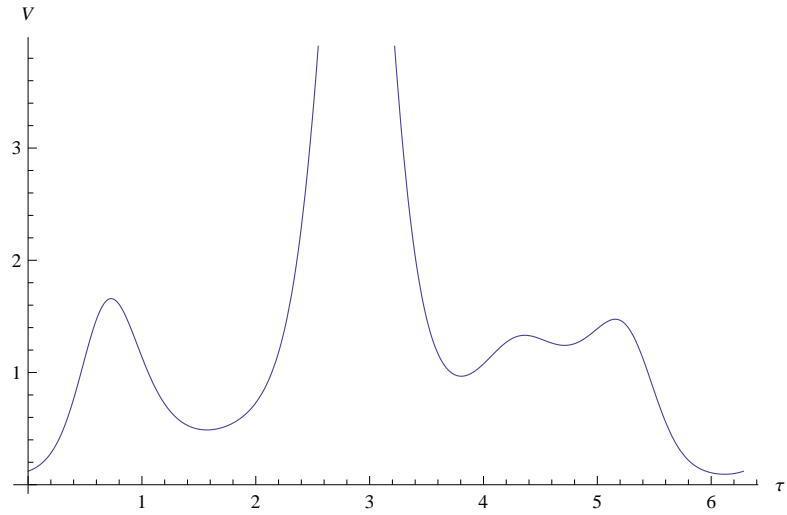}
		\caption{The evolution of the volume for the expression \eqref{KN4.37} with $V_0=1$, $t \in [0,2\pi$].}
	\label{fig:11}
\end{figure}
 The scalar curvature has the form
 \begin{eqnarray}\label{R6}
R&=&2(2+\cos(2\tau))^2\cos^2(3\tau)+(2(2+\cos(2\tau))^2\sin(3\tau)+(6+2\sin(4\tau))\cos(2\tau)+\notag\\& &
+12-4\sin(2\tau)+4\sin(4\tau))\cos(3\tau)+2(2+\cos(2\tau))^2\sin^2(3\tau)+\notag\\& &
+((-6+2\sin(4\tau))\cos(2\tau)+4\sin(4\tau)-4\sin(2\tau)-12)*\notag\\& &
*\sin(3\tau)+2\sin^2(4\tau)+8\cos(4\tau).
\end{eqnarray}
\begin{figure}[h]
	\centering
		\includegraphics[width=0.5 \textwidth]{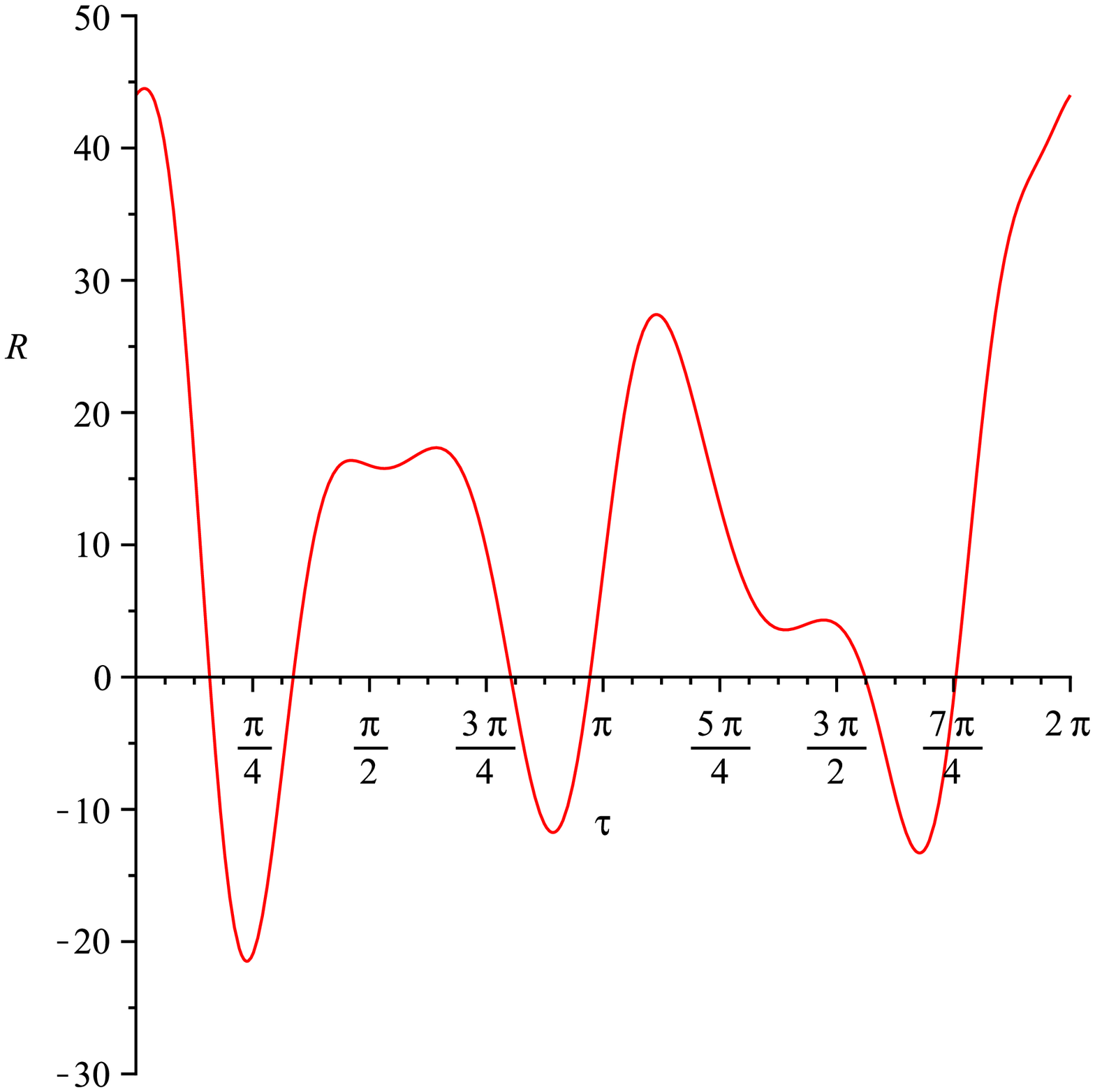}
	\caption{The evolution of the $R$ with respect of the cosmic time $\tau$ for  Eq.\eqref{R6}}
	\label{fig:R4.2.2}
\end{figure}
In Fig.\ref{fig:R4.2.2} we plot the evolution of the $R$ with respect of the cosmic time $\tau$. Again  we have shown that  the Einstein equations admit  the figure-eight knot solution and it  again describe the accelerated and decelerated  expansion phases of the universe.

 \subsection{Example 3.} 
 Now we  present   the figure-eight knot universe induced by the Jacobian elliptic functions. Let the system \eqref{KN2.21}--\eqref{KN2.24}  has the solution 
\begin{eqnarray}\label{KN4.39}
A&=&A_0+[2+\mbox{cn}(2\tau)]\mbox{cn}(3\tau), \\
B&=&B_0+[2+\mbox{cn}(2\tau)]\mbox{sn} (3\tau), \\
C&=&C_0+\mbox{sn}(4\tau).\label{KN4.41}
\end{eqnarray}
\begin{figure}[h]
	\centering
		\includegraphics{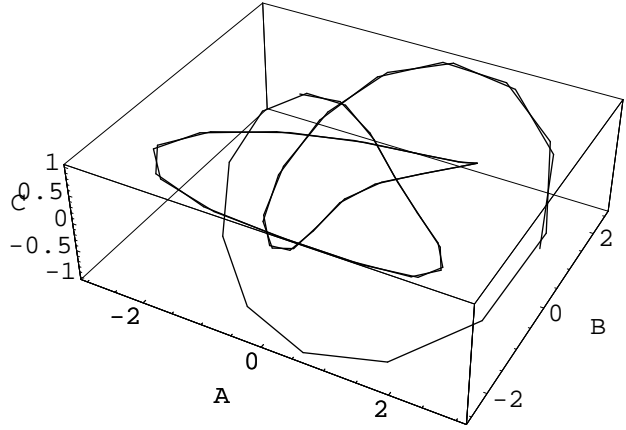}
		\caption{ The knotted closed curve  corresponding to the solution \eqref{KN4.39}--\eqref{KN4.41} with \eqref{KN3.16}, $t \in [0,4\pi$], $k=1/3$.}
	\label{fig:12}
\end{figure}
Note that  $\mbox{cn}(t)$ and $\mbox{sn}(t)$ are  the doubly periodic Jacobian  elliptic functions. Fig.\ref{fig:12} shows the knotted closed curve  corresponding to the solution \eqref{KN4.39}--\eqref{KN4.41} with \eqref{KN3.16}. Substituting the formulas \eqref{KN4.39}--\eqref{KN4.41} into the system \eqref{KN2.21}--\eqref{KN2.24} we get the corresponding expressions for $\rho$ and $p_i$ that gives us the parametric EoS. The evolution of the volume of the universe for \eqref{KN3.16} reads as
  \begin{equation}
V=[2+\mbox{cn}(2\tau)]^2\mbox{cn}(3\tau)\mbox{sn}(3\tau)\mbox{sn}(4\tau).
\end{equation}

 The scalar curvature has the form
 \begin{eqnarray}\label{R7}
R&=&(-18\mbox{sn}(3\tau, k)\mbox{sn}(4\tau, k)k^2(2+\mbox{cn}(2\tau, k))^2\mbox{cn}^3(3\tau, k)+(24(-(3/2)\mbox{sn}(2\tau, k)\mbox{dn}(2\tau, k)\mbox{sn}(4\tau, k)+\notag\\& &
+\mbox{cn}(4\tau, k)\mbox{dn}(4\tau, k)(2+\mbox{cn}(2\tau, k))))(2+\mbox{cn}(2\tau, k))\mbox{dn}(3\tau, k)\mbox{cn}^2(3\tau, k)-\notag\\& &
-(32(-(9/16)\mbox{sn}(4\tau, k)k^2(2+\mbox{cn}(2\tau, k))^2\mbox{sn}^2(3\tau, k)+((\mbox{cn}^2(4\tau, k)k^2+(27/16)\mbox{dn}^2(3\tau, k)+\notag\\& &
+\mbox{dn}^2(4\tau, k)-(1/2)k^2\mbox{sn}^2(2\tau, k)+(1/2)\mbox{dn}^2(2\tau, k))\mbox{cn}^2(2\tau, k)+(-k^2\mbox{sn}^2(2\tau, k)+\notag\\& &
+(27/4)\mbox{dn}^2(3\tau, k)+\mbox{dn}^2(2\tau, k)+4\mbox{dn}^2(4\tau, k)+4\mbox{cn}^2(4\tau, k)k^2)\mbox{cn}(2\tau, k)+4\mbox{dn}^2(4\tau, k)+\notag\\& &
+(27/4)\mbox{dn}^2(3\tau, k)-(1/4)\mbox{dn}^2(2\tau, k)\mbox{sn}^2(2\tau, k)+4\mbox{cn}^2(4\tau, k)k^2)\mbox{sn}(4\tau, k)+\mbox{cn}(4\tau, k)\mbox{dn}(4\tau, k)\notag\\& &
\mbox{dn}(2\tau, k)\mbox{sn}(2\tau, k)(2+\mbox{cn}(2\tau, k))))\mbox{sn}(3\tau, k)\mbox{cn}(3\tau, k)-(24(-(3/2)\mbox{sn}(2\tau, k)\mbox{dn}(2\tau, k)\notag\\& &
\mbox{sn}(4\tau, k)+\mbox{cn}(4\tau, k)\mbox{dn}(4\tau, k)(2+\mbox{cn}(2\tau, k))))(2+\mbox{cn}(2\tau, k))\mbox{sn}^2(3\tau, k)\mbox{dn}(3\tau, k))/(\mbox{cn}(3\tau, k)\notag\\& &
\mbox{sn}(3\tau, k)(2+\mbox{cn}(2\tau, k))^2\mbox{sn}(4\tau, k)).
\end{eqnarray}
\begin{figure}[h]
	\centering
		\includegraphics[width=0.6 \textwidth]{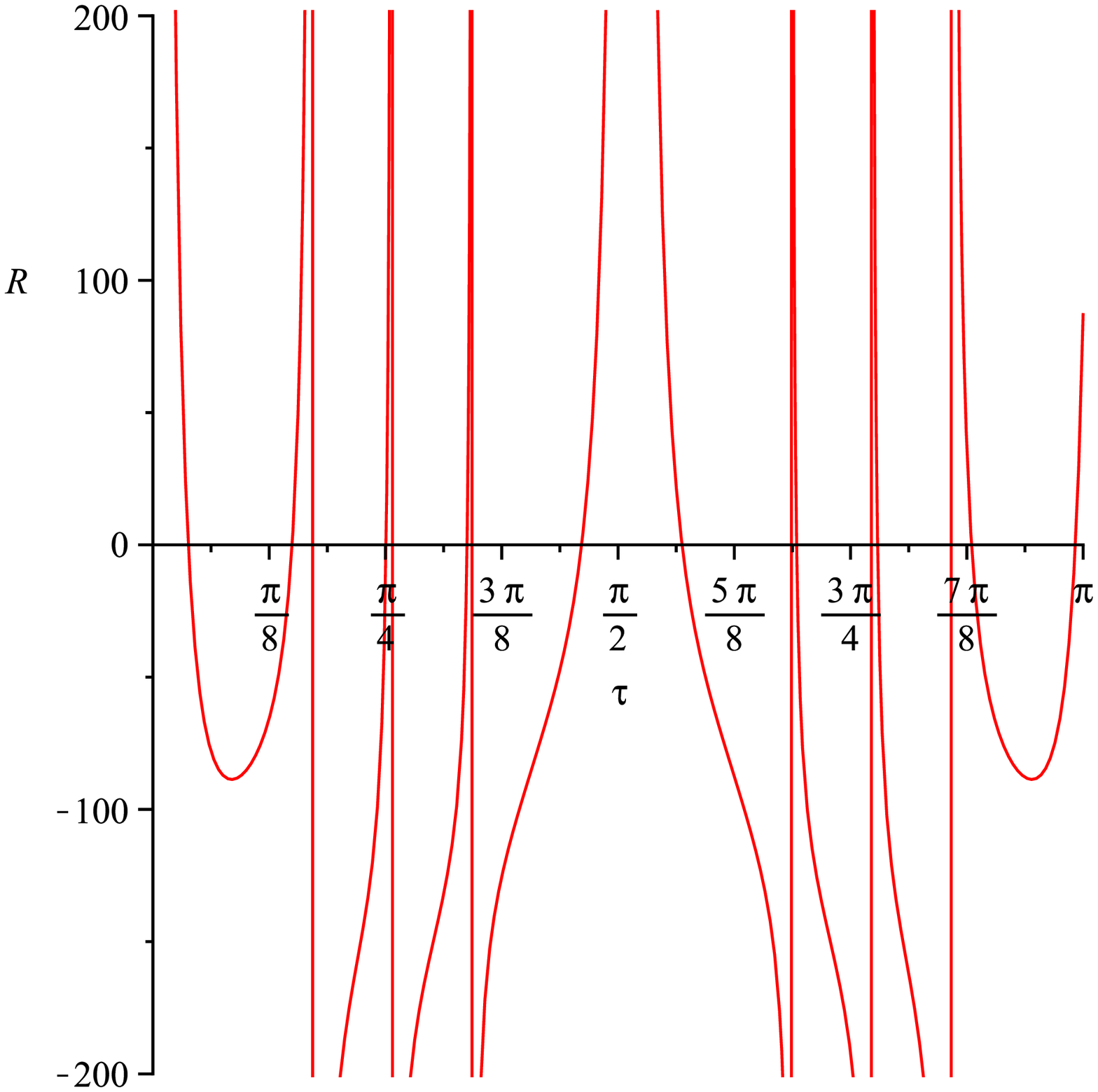}
	\caption{The evolution of the $R$ with respect of the cosmic time $\tau$ for  Eq.\eqref{R7}}
	\label{fig:R4.3.3}
\end{figure}
In Fig.\ref{fig:R4.3.3} we plot the evolution of the $R$ with respect of the cosmic time $\tau$.
     \subsection{Example 4.} We now consider  the following solution of the system \eqref{KN2.25}--\eqref{KN2.28}: 
    \begin{eqnarray}
H_1&=&[2+\mbox{cn}(2\tau)]\mbox{cn}(3\tau), \\
H_2&=&[2+\mbox{cn}(2\tau)]\mbox{sn} (3\tau), \\
H_3&=&\mbox{sn}(4\tau)
    \end{eqnarray}
    which again the trefoil knot universe as shown in Fig.\ref{fig:12} but for the "coordinates" $H_i$. The corresponding parametric EoS reads as \begin{eqnarray}\label{KN4.47}
\rho&=&\frac{D_0}{E_0},\\
p_1&=&-\frac{D_1}{E_1},\\
p_2&=&-\frac{D_2}{E_2},\\
p_3&=&-\frac{D_3}{E_3},\label{KN4.50}
\end{eqnarray}
where
\begin{eqnarray}
D_0&=&(((2+cn(2\tau, k))sn(3\tau, k)+sn(4\tau, k))cn(3\tau, k)+sn(3\tau, k)sn(4\tau, k))\notag\\&
&(2+cn(2\tau, k)),\\
E_0&=&1,\\
D_1&=&2\frac{\partial}{\partial\tau}am(2\tau, k)sn(2\tau, k)sn(3\tau, k)-(3(2+cn(2\tau, k)))cn(3\tau, k)\frac{\partial}{\partial\tau}am(3\tau, k)-\notag\\&
&-4cn(4\tau, k)\frac{\partial}{\partial\tau}am(4\tau, k)-(2+cn(2\tau, k))^2sn(3\tau, k)^2-sn(4\tau, k)^2-\notag\\&
&-(2+cn(2\tau, k))sn(3\tau, k)sn(4\tau, k),\\
E_1&=&1,\\
D_2&=&-4cn(4\tau, k)\frac{\partial}{\partial\tau}am(4\tau, k)+2\frac{\partial}{\partial\tau}am(2\tau, k)sn(2\tau, k)cn(3\tau, k)+(3(2+cn(2\tau, k)))\notag\\&
&\frac{\partial}{\partial\tau}am(3\tau, k)sn(3\tau, k)-sn^2(4\tau, k)-(2+cn(2\tau, k))^2cn^2(3\tau, k)-\notag\\&
&-(2+cn(2\tau, k))cn(3\tau, k)sn(4\tau, k),\\
E_2&=&1,\\
D_3&=&-(2+cn(2\tau, k))^2cn^2(3\tau, k)+(-sn(3\tau, k)cn^2(2\tau, k)+(-4sn(3\tau, k)-\notag\\&
&-3\frac{\partial}{\partial\tau}am(3\tau, k))cn(2\tau, k)+2\frac{\partial}{\partial\tau}am(2\tau, k)sn(2\tau, k)-6\frac{\partial}{\partial\tau}am(3\tau, k)-\notag\\&
&-4sn(3\tau, k))cn(3\tau, k)+3\frac{\partial}{\partial\tau}am(3\tau, k)sn(3\tau, k)cn(2\tau, k)+\notag\\&
&+(6\frac{\partial}{\partial\tau}am(3\tau, k)+2\frac{\partial}{\partial\tau}am(2\tau, k)sn(2\tau, k))sn(3\tau, k)-sn^2(4\tau, k),\\
E_3&=&1.
\end{eqnarray}
\begin{figure}[h]
	\centering
		\includegraphics{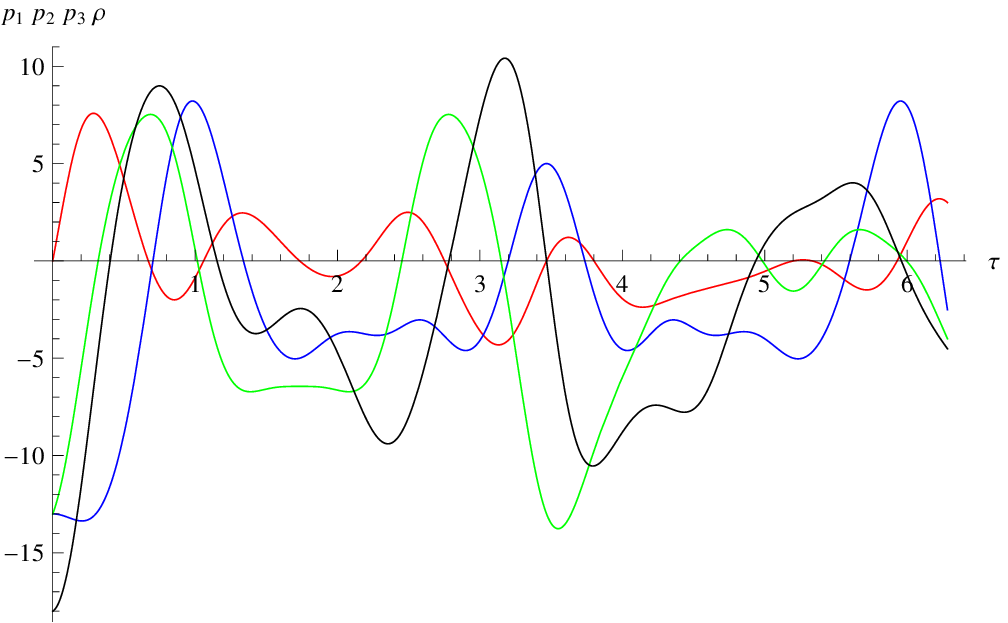}
		\caption{The plot of the EoS \eqref{KN4.47}--\eqref{KN4.50}, $t \in [0,2\pi$], $ k=1/3$, $\rho$(red), $p_1$(blue), $p_2$(green), $p_3$(black).}
	\label{fig:13}
\end{figure}
Its  plot we give in Fig.\ref{fig:13}. 

 The scalar curvature has the form
 \begin{eqnarray}\label{R8}
R&=&2(2+\mbox{cn}(2\tau, k))^2\mbox{cn}(3\tau, k)^2+(2(2+\mbox{cn}(2\tau, k))^2\mbox{sn}(3\tau, k)+(6\mbox{dn}(3\tau, k)+2\mbox{sn}(4\tau, k))\mbox{cn}(2\tau, k)+\notag\\& &
+12\mbox{dn}(3\tau, k)-4\mbox{dn}(2\tau, k)\mbox{sn}(2\tau, k)+4\mbox{sn}(4\tau, k))\mbox{cn}(3\tau, k)+2(2+\mbox{cn}(2\tau, k))^2\mbox{sn}(3\tau, k)^2+\notag\\& &
+((-6\mbox{dn}(3\tau, k)+2\mbox{sn}(4\tau, k))\mbox{cn}(2\tau, k)+4\mbox{sn}(4\tau, k)-4\mbox{dn}(2\tau, k)\mbox{sn}(2\tau, k)-12\mbox{dn}(3\tau, k))*\notag\\& &
*\mbox{sn}(3\tau, k)+2\mbox{sn}(4\tau, k)^2+8\mbox{cn}(4\tau, k)\mbox{dn}(4\tau, k).
\end{eqnarray}
\begin{figure}[h]
	\centering
		\includegraphics[width=0.5 \textwidth]{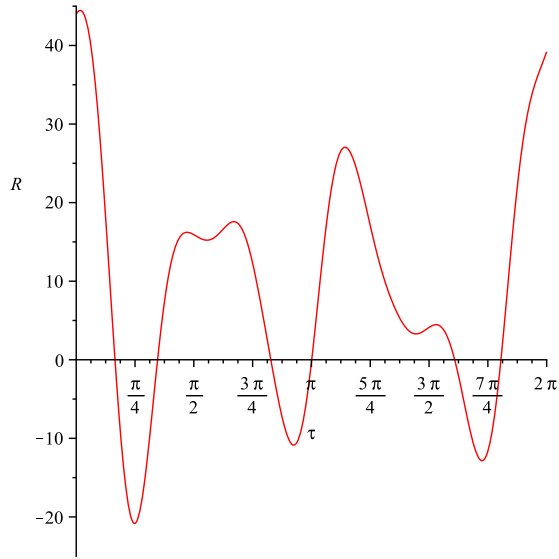}
	\caption{The evolution of the $R$ with respect of the cosmic time $\tau$ for  Eq.\eqref{R8}}
	\label{fig:R4.4.4}
\end{figure}
In Fig.\ref{fig:R4.4.4} we plot the evolution of the $R$ with respect of the cosmic time $\tau$.
\section{Other unknotted models of the universe}
In this section we would like to present some unknotted but closed curve solutions of the Einstein equation for the Bianchi I  type metric. As an examples we consider  the spiky and Mobious strip universe solutions. 
\subsection{Spiky universe  solutions}
Our aim in  this subsection is  to present some unknotted closed curve solutions  namely the spiky  universe solutions. 
\subsubsection{Example 1} 
\begin{figure}[h]
	\centering
		\includegraphics{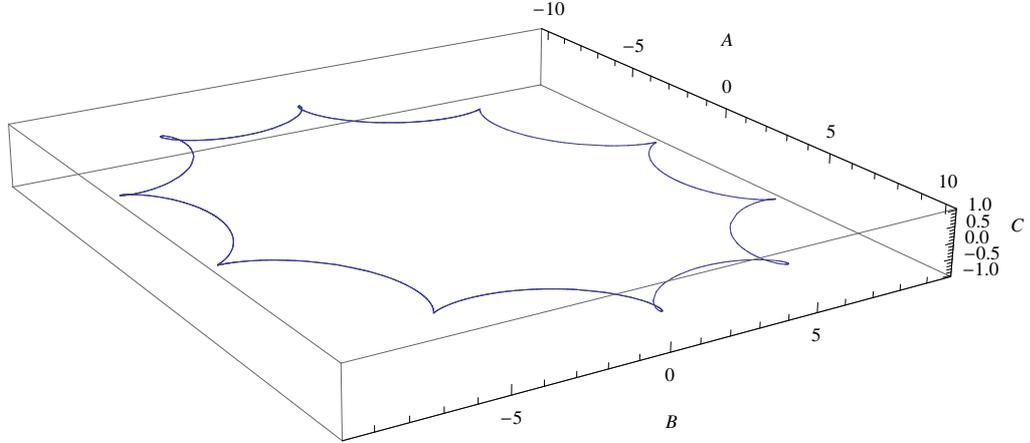}
	\caption{The spiky universe for \eqref{KN5.13}--\eqref{KN5.15}, $n=10, \alpha=1$.}
	\label{fig:14}
\end{figure}

\begin{figure}[h]
	\centering
		\includegraphics{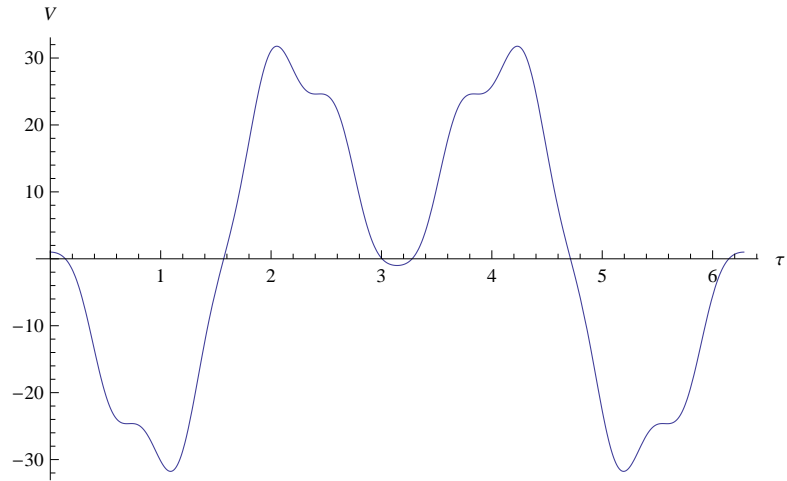}
	\caption{ The evolution of the volume for  (5.16), $n=10$, $\alpha=1$.}
	\label{fig:15}
\end{figure}

Let  our universe is filled by the fluid with the following parametric EoS
\begin{eqnarray}
\rho&=&\frac{D_8}{E_8},\\
p_1&=&-\frac{D_9}{E_9},\\
p_2&=&-\frac{D_{10}}{E_{10}},\\
p_3&=&-\frac{D_{11}}{E_{11}},
\end{eqnarray}
where
\begin{eqnarray}
D_8&=&-[\alpha\sin((n-1)\tau)(n-1)+\alpha(n-1)\sin(\tau)][\alpha\cos((n-1)\tau)(n-1)-\notag\\
& &\alpha(n-1)\cos(\tau)]\sin(\tau)+[\alpha\cos((n-1)\tau)(n-1)-\alpha(n-1)\cos(\tau)]\cos(\tau)\times\notag\\
& &[\alpha\cos((n-1)\tau)+\alpha(n-1)\cos(\tau)]-[\alpha\sin((n-1)\tau)(n-1)+\alpha(n-1)\sin(\tau)]\times\notag\\
& &[\alpha\sin((n-1)\tau)-\alpha(n-1)\sin(\tau)]\cos(\tau),\\
E_8&=&[\alpha\cos((n-1)\tau)+\alpha(n-1)\cos(\tau)][\alpha\sin((n-1)\tau)-\alpha(n-1)\sin(\tau)]\sin(\tau),\\
D_9&=&[-\alpha\sin((n-1)\tau)(n-1)^2+\alpha(n-1)\sin(\tau)]\sin(\tau)-[\alpha\sin((n-1)\tau)-\alpha(n-1)\sin(\tau)]\sin(\tau)+\notag\\
& &[\alpha\cos((n-1)\tau)(n-1)-\alpha(n-1)\cos(\tau)]\cos(\tau),\\
E_9&=&[\alpha\sin((n-1)\tau)-\alpha(n-1)\sin(\tau)]\sin(\tau),\\
D_{10}&=&[\alpha\cos((n-1)\tau)+\alpha(n-1)\cos(\tau)]\sin(\tau)+\sin(\tau)[\alpha\cos((n-1)\tau)(n-1)^2+\alpha(n-1)\cos(\tau)]+\notag\\
& &[\alpha\sin((n-1)\tau)(n-1)+\alpha(n-1)\sin(\tau)]\cos(\tau),\\
E_{10}&=&-[\alpha\cos((n-1)\tau)+\alpha(n-1)\cos(\tau)]\sin(\tau),\\
D_{11}&=&[\alpha\sin((n-1)\tau)-\alpha(n-1)\sin(\tau)][-\alpha\cos((n-1)\tau)(n-1)^2-\alpha(n-1)\cos(\tau)]+\notag\\
& &[\alpha\cos((n-1)\tau)+\alpha(n-1)\cos(\tau)][-\alpha\sin((n-1)\tau)(n-1)^2+\alpha(n-1)\sin(\tau)]-\notag\\
& &[\alpha\sin((n-1)\tau)(n-1)+\alpha(n-1)\sin(\tau)][\alpha\cos((n-1)\tau)(n-1)-\alpha(n-1)\cos(\tau)],\\
E_{11}&=&[\alpha\cos((n-1)\tau)+\alpha(n-1)\cos(\tau)][\alpha\sin((n-1)\tau)-\alpha(n-1)\sin(\tau)].
\end{eqnarray}
Substituting these expressions for the pressuries and the density of energy into the system \eqref{KN2.21}--\eqref{KN2.24}, we obtain the following its solution 
\begin{eqnarray}\label{KN5.13}
A&=&\alpha\cos[(n-1)\tau]+\alpha(n-1)\cos[\tau], \\
B&=&\alpha\sin[(n-1)\tau]-\alpha(n-1)\sin[\tau], \\
C&=&\sin (\tau).\label{KN5.15}
    \end{eqnarray}
 It is the  spiky like solution so that such solutions we call the spike universe. Its plot presented in Fig.\ref{fig:14} for the initial conditions $A(0)=\alpha n=10, B(0)=0, C(0)=0$. 
  Let us  calculate the volume of this  universe. It is given by
  \begin{equation}\label{KN5.16}
V=\alpha^2[\cos[(n-1)\tau]+(n-1)\cos[\tau]][\sin[(n-1)\tau]-(n-1)\sin[\tau]]\sin (\tau).
\end{equation}
In Fig.\ref{fig:15} shown  the evolution of the volume for  \eqref{KN5.16}, $n=10, \alpha=1$.  
 The scalar curvature has the form
 \begin{eqnarray}\label{R9}
R&=&(-2\cos(\tau)(n-1)\cos^2((n-1)\tau)+((6(4/3-2n+n^2))\sin(\tau)\sin((n-1)\tau)-\notag\\& &
-(2((n-2)\cos(\tau)^2+\sin^2(\tau)(n^2-3n+4)))(n-1))\cos((n-1)\tau)+2\cos(\tau)(\sin^2((n-1)\tau)+\notag\\& &
+\sin(\tau)(n^2-4n+6)\sin((n-1)\tau)+(\cos(\tau)^2-5\sin^2(\tau))(n-1))(n-1))/((\cos((n-1)\tau)+\notag\\& &
+\cos(\tau)(n-1))(-\sin((n-1)\tau)+(n-1)\sin(\tau))\sin(\tau)).
\end{eqnarray}
\begin{figure}[h]
	\centering
		\includegraphics[width=0.5 \textwidth]{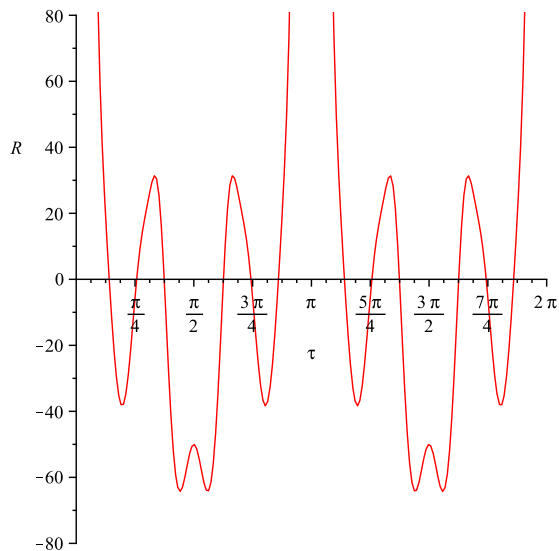}
	\caption{The evolution of the $R$ with respect of the cosmic time $\tau$ for  Eq.\eqref{R9}}
	\label{fig:R5.1.1}
\end{figure}
In Fig.\ref{fig:R5.1.1} we plot the evolution of the $R$ with respect of the cosmic time $\tau$. In this example,  we have shown that  the Einstein equations admit  the spike-like  solution. We can show this solution  describes the accelerated and decelerated  expansion phases of the universe.
\subsubsection{Example 2}
The system \eqref{KN2.25}--\eqref{KN2.28} admits the following solution
\begin{eqnarray}
H_1&=&\alpha\cos[(n-1)\tau]+\alpha(n-1)\cos[\tau], \\
H_2&=&\alpha\sin[(n-1)\tau]-\alpha(n-1)\sin[\tau], \\
H_3&=&\sin (\tau).
    \end{eqnarray}
    The corresponding EoS takes the form
    \begin{eqnarray}
\rho&=&\frac{D_{12}}{E_{12}},\\
p_1&=&-\frac{D_{13}}{E_{13}},\\
p_2&=&-\frac{D_{14}}{E_{14}},\\
p_3&=&-\frac{D_{15}}{E_{15}},
\end{eqnarray}
where
\begin{eqnarray}
D_{12}&=&[\alpha\cos((n-1)\tau)+\alpha(n-1)\cos(\tau)][\alpha\sin((n-1)\tau)+[1-\alpha(n-1)]\sin(\tau)]+\notag\\
& &[\alpha\sin((n-1)\tau)-\alpha(n-1)\sin(\tau)]\sin(\tau),\\
E_{12}&=&1,\\
D_{13}&=&\alpha(n-1)[\cos((n-1)\tau)-\cos(\tau)]+\cos(\tau)+[\alpha\sin((n-1)\tau)-\alpha(n-1)\sin(\tau)]^2+\notag\\
& &[\alpha\sin((n-1)\tau)+[1-\alpha(n-1)]\sin(\tau)]\sin(\tau),\\
E_{13}&=&1,\\
D_{14}&=&-\alpha\sin((n-1)\tau)(n-1)-\alpha(n-1)\sin(\tau)+\cos(\tau)+[\alpha\cos((n-1)\tau)+\alpha(n-1)\cos(\tau)]^2+\notag\\
& &\sin(\tau)^2+[\alpha\cos((n-1)\tau)+\alpha*(n-1)\cos(\tau)]\sin(\tau),\\
E_{14}&=&1,\\
D_{15}&=&\alpha(n-1)[\cos((n-1)\tau)-\cos(\tau)-\sin((n-1)\tau)-\sin(\tau)]+\notag\\
& &[\alpha\sin((n-1)\tau)-\alpha(n-1)\sin(\tau)]^2+[\alpha\cos((n-1)\tau)+\alpha(n-1)\cos(\tau)]^2+\notag\\
& &[\alpha\cos((n-1)\tau)+\alpha(n-1)\cos(\tau)][\alpha\sin((n-1)\tau)-\alpha(n-1)\sin(\tau)],\\
E_{15}&=&1.
\end{eqnarray}

 The scalar curvature has the form
 \begin{eqnarray}\label{R10}
R&=&2\alpha^2\cos((n-1)\tau)^2+4\alpha((1/2)\alpha\sin((n-1)\tau)+(1/2+(-(1/2)n+1/2)\alpha)\sin(\tau)+\notag\\& &
+(n-1)(\alpha\cos(\tau)+1/2))\cos((n-1)\tau)+2\alpha^2\sin((n-1)\tau)^2+2\alpha((1+(2-2n)\alpha)\sin(\tau)+\notag\\& &
+(\alpha\cos(\tau)-1)(n-1))\sin((n-1)\tau)+(2+2\alpha^2(n-1)^2+(2-2n)\alpha)\sin(\tau)^2-(2(n-1))\alpha*\notag\\& &
*(1+(-1+\alpha(n-1))\cos(\tau))\sin(\tau)+(2(\alpha^2(n-1)^2\cos(\tau)+1+\alpha(-n+1)))\cos(\tau).
\end{eqnarray}
\begin{figure}[h]
	\centering
		\includegraphics[width=0.5 \textwidth]{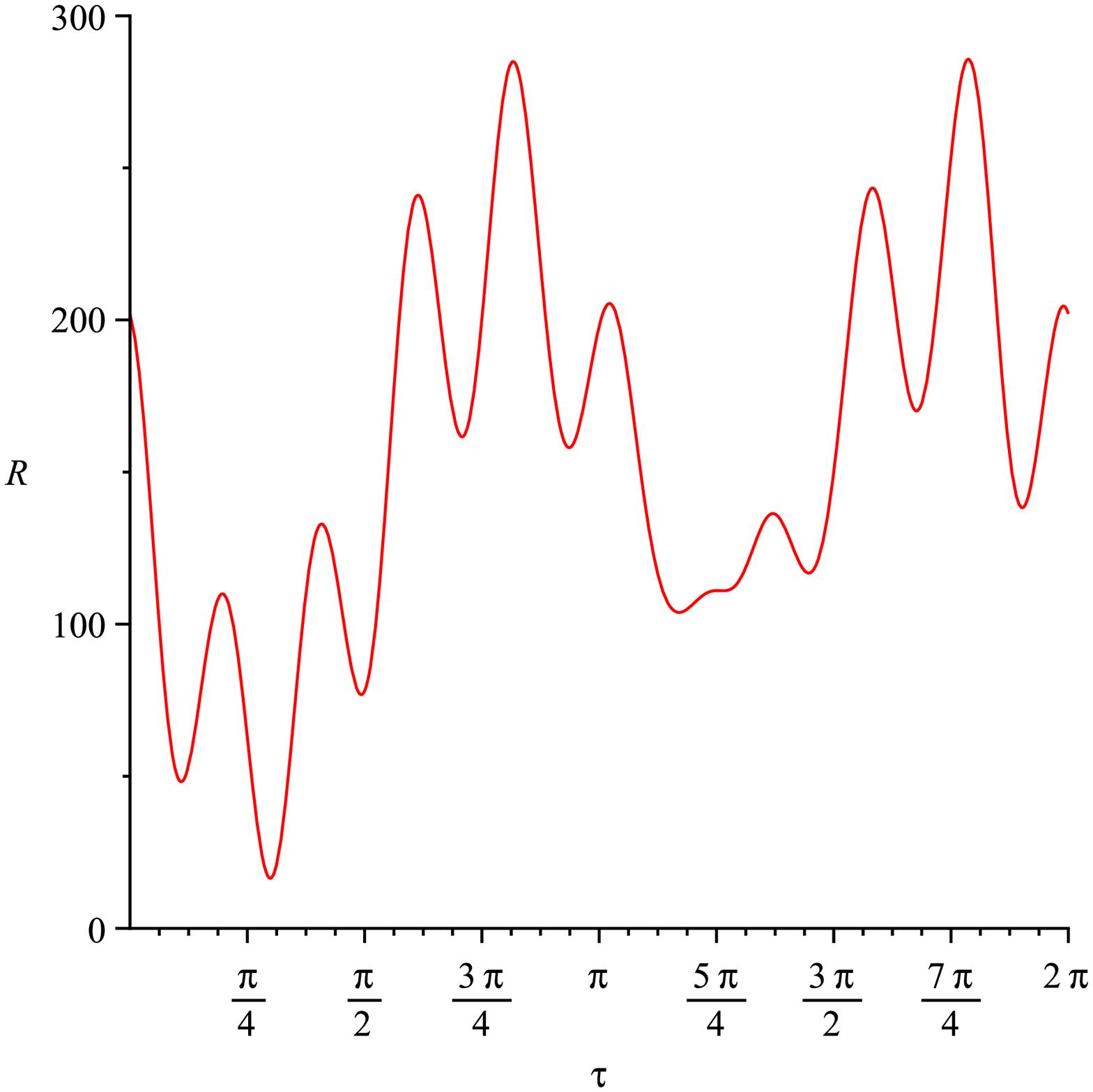}
	\caption{The evolution of the $R$ with respect of the cosmic time $\tau$ for  Eq.\eqref{R10}}
	\label{fig:R5.1.2}
\end{figure}
In Fig.\ref{fig:R5.1.2} we plot the evolution of the $R$ with respect of the cosmic time $\tau$.
\subsubsection{Example 3}
Our next solution for the system \eqref{KN2.25}--\eqref{KN2.28} is given by
\begin{eqnarray}
H_1&=&\alpha\cos[(n-1)\tau]-\alpha(n-1)\cos[\tau], \\
H_2&=&\alpha\sin[(n-1)\tau]-\alpha(n-1)\sin[\tau], \\
H_3&=&\sin (\tau).
\end{eqnarray}
   In Fig.27 we plot this spiky type solution.  
   \begin{figure}[h]
	\centering
		\includegraphics[width=0.5 \textwidth]{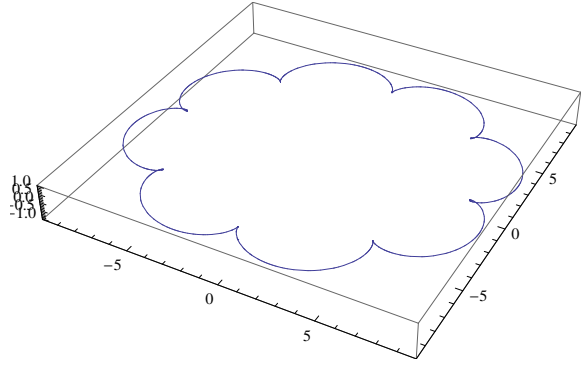}
	\caption{The evolution of the spiky type solution (5.34)-(5.36) with $n=10, \alpha=1$.}
	\label{fig:R5.1.3}
\end{figure}
    The corresponding EoS takes the form
    \begin{eqnarray}
\rho&=&\frac{D_{16}}{E_{16}},\\
p_1&=&-\frac{D_{17}}{E_{17}},\\
p_2&=&-\frac{D_{18}}{E_{18}},\\
p_3&=&-\frac{D_{19}}{E_{19}},
\end{eqnarray}
where
\begin{eqnarray}
D_{16}&=&[\alpha\cos((n-1)\tau)-\alpha(n-1)\cos(\tau)][\alpha\sin((n-1)\tau)+[1-\alpha(n-1)]\sin(\tau)]+\notag\\
& &[\alpha\sin((n-1)\tau)-\alpha(n-1)\sin(\tau)]\sin(\tau),\\
E_{16}&=&1,\\
D_{17}&=&\alpha(n-1)[\cos((n-1)\tau)-\cos(\tau)]+\cos(\tau)+[\alpha\sin((n-1)\tau)-\alpha(n-1)\sin(\tau)]^2+\notag\\
& &[\alpha\sin((n-1)\tau)+[1-\alpha(n-1)]\sin(\tau)]\sin(\tau),\\
E_{17}&=&1,\\
D_{18}&=&-\alpha\sin((n-1)\tau)(n-1)+\alpha(n-1)\sin(\tau)+\cos(\tau)+[\alpha\cos((n-1)\tau)-\alpha(n-1)\cos(\tau)]^2+\notag\\
& &\sin(\tau)^2+[\alpha\cos((n-1)\tau)-\alpha*(n-1)\cos(\tau)]\sin(\tau),\\
E_{18}&=&1,\\
D_{19}&=&\alpha(n-1)[\cos((n-1)\tau)-\cos(\tau)-\sin((n-1)\tau)+\sin(\tau)]+\notag\\
& &[\alpha\sin((n-1)\tau)-\alpha(n-1)\sin(\tau)]^2+[\alpha\cos((n-1)\tau)-\alpha(n-1)\cos(\tau)]^2+\notag\\
& &[\alpha\cos((n-1)\tau)-\alpha(n-1)\cos(\tau)][\alpha\sin((n-1)\tau)-\alpha(n-1)\sin(\tau)],\\
E_{19}&=&1.
\end{eqnarray}  
 The scalar curvature has the form
 \begin{eqnarray}\label{R11}
R&=&2\alpha^2\cos((n-1)\tau)^2-(4(-(1/2)\alpha\sin((n-1)\tau)+(-1/2+((1/2)n-1/2)\alpha)\sin(\tau)+\notag\\& &
+(n-1)(-1/2+\alpha\cos(\tau))))\alpha\cos((n-1)\tau)+2\alpha^2\sin((n-1)\tau)^2-\notag\\& &
-(2((-1+(-2+2n)\alpha)\sin(\tau)+(\alpha\cos(\tau)+1)(n-1)))\alpha\sin((n-1)\tau)+\notag\\& &
+(2+2\alpha^2(n-1)^2+(-2n+2)\alpha)\sin(\tau)^2+(2(n-1))(1+(-1+\alpha(n-1))*\notag\\& &
*\cos(\tau))\alpha\sin(\tau)+2\cos(\tau)(\alpha^2(n-1)^2\cos(\tau)+1+(1-n)\alpha).
\end{eqnarray}
\begin{figure}[h]
	\centering
		\includegraphics[width=0.5 \textwidth]{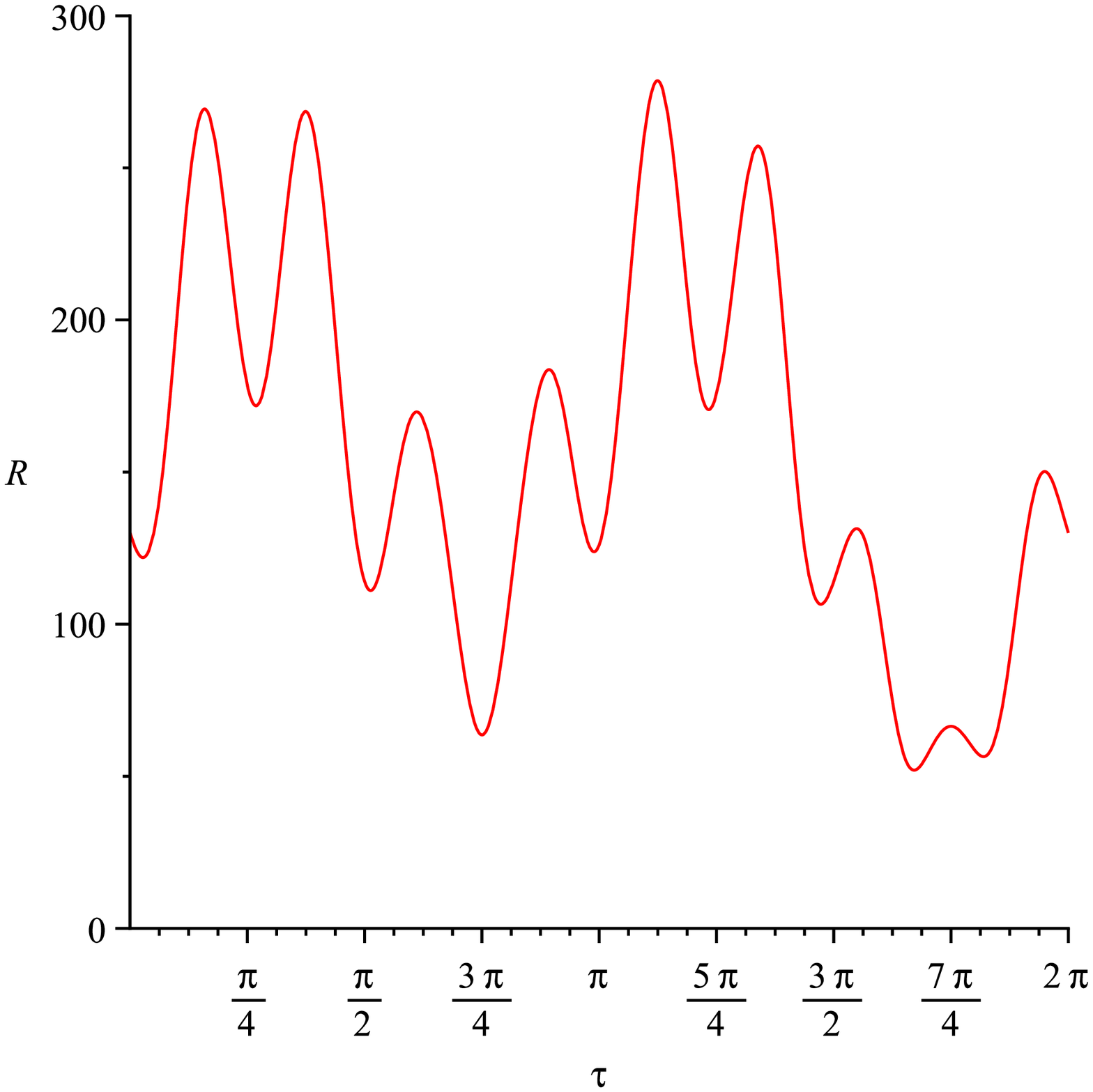}
	\caption{The evolution of the $R$ with respect of the cosmic time $\tau$ for  Eq.\eqref{R11}}
	\label{fig:R5.1.3}
\end{figure}
In Fig.\ref{fig:R5.1.3} we plot the evolution of the $R$ with respect of the cosmic time $\tau$.  
\subsection{M$\ddot{o}$bius strip universe solutions}
    If we consider the model with the "cosmological constant", then the systems \eqref{KN2.21}--\eqref{KN2.24} and \eqref{KN2.25}--\eqref{KN2.28} take the form, respectively
    \begin{eqnarray}\label{KN5.50}
\frac{\dot{A}\dot{B}}{AB}+\frac{\dot{B}\dot{C}}{BC}+\frac{\dot{C}\dot{A}}{CA}-\rho-\Lambda&=&0,\\
\frac{\ddot{B}}{B}+\frac{\ddot{C}}{C}+\frac{\dot{B}\dot{C}}{BC}+p_1-\Lambda&=&0,\\
\frac{\ddot{C}}{C}+\frac{\ddot{A}}{A}+\frac{\dot{C}\dot{A}}{CA}+p_2-\Lambda&=&0,\\
\frac{\ddot{A}}{A}+\frac{\ddot{B}}{B}+\frac{\dot{A}\dot{B}}{AB}+p_3-\Lambda&=&0\label{KN5.53}
\end{eqnarray}
and 
\begin{eqnarray}\label{KN5.54}
H_1H_2+H_2H_3+H_1H_3-\rho-\Lambda&=&0,\\
\dot{H}_2+
\dot{H}_3+H^2_2+H^2_3+H_2H_3+p_1-\Lambda&=&0,\\
\dot{H}_3+\dot{H}_1+H^2_3+H^2_1+H_3H_1+p_2-\Lambda&=&0,\\
\dot{H}_1+
\dot{H}_2+H^2_1+H^2_2+H_1H_2+p_3-\Lambda&=&0.\label{KN5.57}
\end{eqnarray}Now we want to present some  solutions of these systems. Consider examples. 
\subsubsection{Example 1}
\begin{figure}[h]
	\centering
		\includegraphics{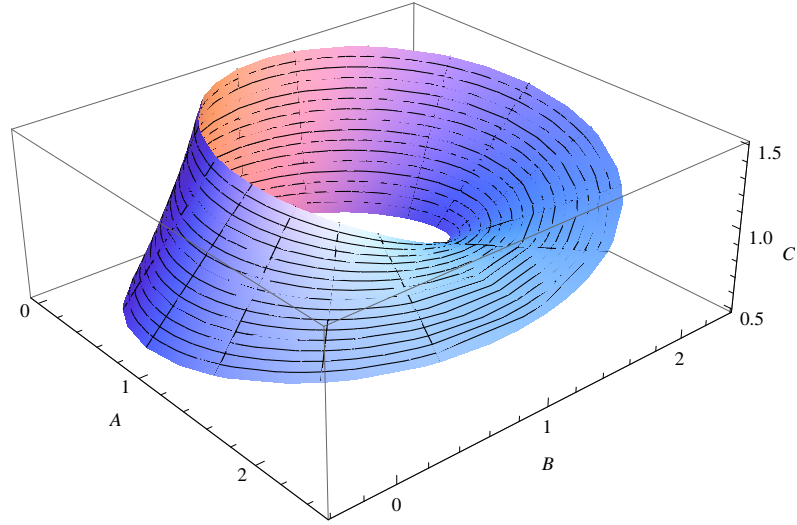}
	\caption{The plot of the M$\ddot{o}$bius strip universe for \eqref{KN5.58}--\eqref{KN5.60} with \eqref{KN3.16} and  $\tau=0\rightarrow 2\pi$ and  $\Lambda=[-1. 1]$}
	\label{fig:16}
\end{figure}

\begin{figure}[h]
	\centering
		\includegraphics{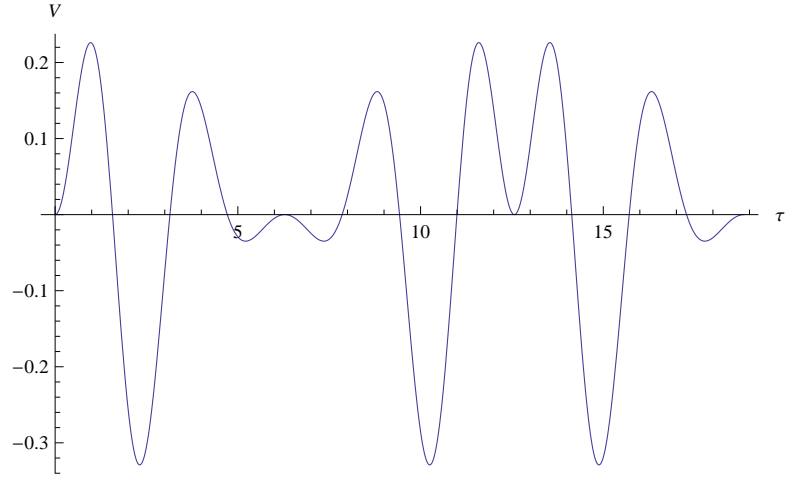}
	\caption{The  evolution of the volume of the M$\ddot{o}$bius strip universe for \eqref{KN5.58}--\eqref{KN5.60} with \eqref{KN3.16} and $\alpha=\Lambda =1$.}
	\label{fig:17}
\end{figure}

 One of the simplest solutions of \eqref{KN5.50}--\eqref{KN5.53} is given by
    \begin{eqnarray}\label{KN5.58}
A&=&A_0+\left(1+\frac{1}{2}\Lambda \cos \frac{\tau}{2}\right)\cos \tau, \\
B&=&B_0+\left(1+\frac{1}{2}\Lambda\cos\frac{\tau}{2}\right)\sin\tau,\\
C&=&C_0+\frac{1}{2}\Lambda\sin \frac{\tau}{2}.\label{KN5.60}
    \end{eqnarray}
    It is the parametric equation of the M$\ddot{o}$bius strip and, hence, such model we call the M$\ddot{o}$bius strip universe. Its plot was presented in Fig.\ref{fig:16}.
The evolution of the volume of the M$\ddot{o}$bius strip universe for \eqref{KN5.58}--\eqref{KN5.60} with \eqref{KN3.16} reads as
  \begin{equation}
V=0.5\Lambda\left(1+\frac{1}{2}\Lambda \cos \frac{\tau}{2}\right)^2\cos \tau\sin\tau\sin \frac{\tau}{2}.
\end{equation}
The  evolution of the volume with\eqref{KN3.16} and $\alpha=\Lambda =1$  presented in Fig.\ref{fig:17}.

    The corresponding EoS takes the form
    \begin{eqnarray}
\rho&=&\frac{D_{20}}{E_{20}},\\
p_1&=&-\frac{D_{21}}{E_{21}},\\
p_2&=&-\frac{D_{22}}{E_{22}},\\
p_3&=&-\frac{D_{23}}{E_{23}},
\end{eqnarray}
where
\begin{eqnarray}
D_{20}&=&[\frac{1}{4}\Lambda\sin(\frac{\tau}{2})\cos(\tau)+(1+\frac{1}{2}\Lambda\cos(\frac{\tau}{2}))\sin(\tau)][\frac{1}{4}\Lambda\sin(\frac{\tau}{2})\sin(\tau)-(1+\frac{1}{2}\Lambda\cos(\frac{\tau}{2})\cos(\tau)]\times\notag\\& &[C_0+\frac{1}{2}\Lambda\sin(\frac{\tau}{2})]+\frac{\Lambda}{4}[-\frac{1}{4}\Lambda\sin(\frac{\tau}{2})\sin(\tau)+(1+\frac{1}{2}\Lambda\cos(\frac{\tau}{2}))\cos(\tau)]\cos(\frac{\tau}{2})\times\notag\\& &[A_0+(1+\frac{1}{2})\Lambda\cos(\frac{\tau}{2})\cos(\tau)]+\frac{\Lambda}{4}[-\frac{1}{4}\Lambda\sin(\frac{\tau}{2})\cos(\tau)-(1+\frac{1}{2}\Lambda\cos(\frac{\tau}{2}))\sin(\tau)]\cos(\frac{\tau}{2})\times\notag\\& &[B_0+(1+\frac{1}{2})\Lambda\cos(\frac{\tau}{2}))\sin(\tau)]-\Lambda[A_0+(1+\frac{1}{2})\Lambda\cos(\frac{\tau}{2}))\cos(\tau)]\times\notag\\& &[B_0+(1+\frac{1}{2}\Lambda\cos(\frac{\tau}{2}))\sin(\tau)][C_0+\frac{1}{2}\Lambda\sin(\frac{\tau}{2})],\\
E_{20}&=&[A_0+(1+\frac{1}{2}\Lambda\cos(\frac{\tau}{2}))\cos(\tau)][B_0+(1+\frac{1}{2}\Lambda\cos(\frac{\tau}{2}))\sin(\tau)][C_0+\frac{1}{2}\Lambda\sin(\frac{\tau}{2})],\\
D_{21}&=&[C_0+\frac{1}{2}\Lambda\sin(\frac{\tau}{2})][-\frac{1}{8}\sin(\tau)\Lambda\cos(\frac{\tau}{2})-\frac{1}{2}\Lambda\sin(\frac{\tau}{2})\cos(\tau)-(1+\frac{1}{2}\Lambda\cos(\frac{\tau}{2}))\sin(\tau)]-\notag\\& &\frac{\Lambda}{8}[B_0+(1+\frac{1}{2}\Lambda\cos(\frac{\tau}{2}))\sin(\tau)]\sin(\frac{\tau}{2})+\frac{\Lambda}{4}[-\frac{1}{4}\Lambda\sin(\frac{\tau}{2})\sin(\tau)+(1+\frac{1}{2}\Lambda\cos(\frac{\tau}{2}))\times\notag\\& &\cos(\tau)]\cos(\frac{\tau}{2})-\Lambda[B_0+(1+\frac{1}{2}\Lambda\cos(\frac{\tau}{2}))\sin(\tau)][C_0+\frac{1}{2}\Lambda\sin(\frac{\tau}{2})],\\
E_{21}&=&[B_0+(1+\frac{1}{2}\Lambda\cos(\frac{\tau}{2}))\sin(\tau)][C_0+\frac{1}{2}\Lambda\sin(\frac{\tau}{2})],\\
D_{22}&=&-\frac{\Lambda}{8}[A_0+(1+\frac{1}{2}\cos(\frac{\tau}{2}))\cos(\tau)]\Lambda\sin(\frac{\tau}{2})+[C_0+\frac{1}{2}\Lambda\sin(\frac{\tau}{2})][-\frac{\Lambda}{8}\cos(\tau)\cos(\frac{\tau}{2})+\notag\\& &\frac{1}{2}\Lambda\sin(\frac{\tau}{2})\sin(\tau)-(1+\frac{1}{2}\Lambda\cos(\frac{\tau}{2}))\cos(\tau)]+\frac{1}{4}[-\frac{1}{4}\Lambda\sin(\frac{\tau}{2})\cos(\tau)-\notag\\& &(1+\frac{1}{2}\Lambda\cos(\frac{\tau}{2}))\sin(\tau)]\Lambda\cos(\frac{\tau}{2})-\Lambda[A_0+(1+\frac{1}{2}\Lambda\cos(\frac{\tau}{2}))\cos(\tau)][C_0+\frac{1}{2}\Lambda\sin(\frac{\tau}{2})],\\
E_{22}&=&[A
_0+(1+\frac{1}{2}\Lambda\cos(\frac{\tau}{2}))\cos(\tau)][C_0+\frac{1}{2}\Lambda\sin(\frac{\tau}{2})],\\
D_{23}&=&[B_0+(1+\frac{1}{2}\Lambda\cos(\frac{\tau}{2}))\sin(\tau)][-\frac{\Lambda}{8}\cos(\tau)\cos(\frac{\tau}{2})+\frac{1}{2}\Lambda\sin(\frac{\tau}{2})\sin(\tau)-(1+\frac{1}{2}\Lambda\cos(\frac{\tau}{2}))\times\notag\\& &\cos(\tau)]+[A_0+(1+\frac{1}{2}\Lambda\cos(\frac{\tau}{2}))\cos(\tau)][-\frac{\Lambda}{8}\sin(\tau)\cos(\frac{\tau}{2})-\frac{1}{2}\Lambda\sin(\frac{\tau}{2})\cos(\tau)-\notag\\& &(1+\frac{1}{2}\Lambda\cos(\frac{\tau}{2}))\sin(\tau)]-[\frac{1}{4}\Lambda\sin(\frac{\tau}{2})\cos(\tau)+(1+\frac{1}{2}\Lambda\cos(\frac{\tau}{2}))\sin(\tau)]\times\notag\\& &[-\frac{1}{4}\Lambda\sin(\frac{\tau}{2})\sin(\tau)+(1+\frac{1}{2}\Lambda\cos(\frac{\tau}{2}))\cos(\tau)]-\notag\\& &\Lambda[A_0+(1+\frac{1}{2}\Lambda\cos(\frac{\tau}{2}))\cos(\tau)][B_0+(1+\frac{1}{2}\Lambda\cos(\frac{\tau}{2}))\sin(\tau)],\\
E_{23}&=&[B_0+(1+\frac{1}{2}\Lambda\cos(\frac{\tau}{2}))\sin(\tau)][A_0+(1+\frac{1}{2}\Lambda\cos(\frac{\tau}{2}))\cos(\tau)].
\end{eqnarray}
 The scalar curvature has the form
 \begin{eqnarray}\label{R12}
R&=&((-2\sin(\tau)^2\Lambda^2+2\cos(\tau)^2\Lambda^2)\cos((1/2)\tau)^3+\notag\\& &
+(-8\sin(\tau)^2\Lambda-17\cos(\tau)\sin(\tau)\sin((1/2)\tau)\Lambda^2+\notag\\& &
+8\cos(\tau)^2\Lambda)\cos((1/2)\tau)^2+((6\sin(\tau)^2\Lambda^2-6\cos(\tau)^2\Lambda^2)\sin((1/2)\tau)^2-\notag\\& &
-60\cos(\tau)\sin(\tau)\sin((1/2)\tau)\Lambda+8\cos(\tau)^2-8\sin(\tau)^2)\cos((1/2)\tau)+\notag\\& &
+(\sin((1/2)\tau)^2\Lambda^2\cos(\tau)\sin(\tau)+(12\sin(\tau)^2\Lambda-12\cos(\tau)^2\Lambda)\sin((1/2)\tau)-\notag\\& &
-52\cos(\tau)\sin(\tau))\sin((1/2)\tau))/(\sin(\tau)\cos(\tau)(2+\Lambda\cos((1/2)\tau))^2\sin((1/2)\tau)).
\end{eqnarray}
\begin{figure}[h]
	\centering
		\includegraphics[width=0.5 \textwidth]{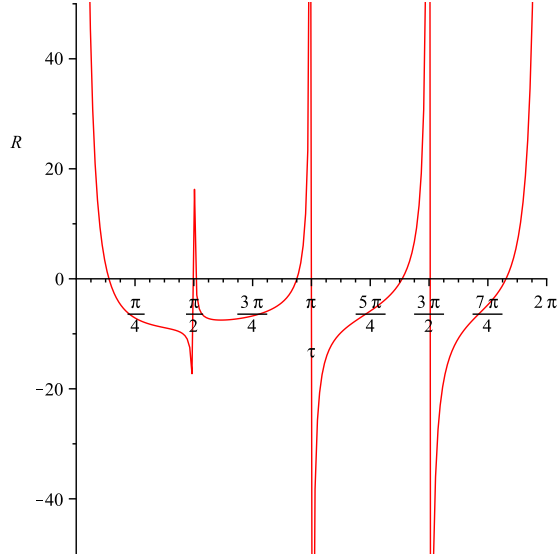}
	\caption{The evolution of the $R$ with respect of the cosmic time $\tau$ for  Eq.\eqref{R12}}
	\label{fig:R5.2.1}
\end{figure}
In Fig.\ref{fig:R5.2.1} we plot the evolution of the $R$ with respect of the cosmic time $\tau$. In this subsubsection,  we have shown that  the Einstein equations have the M$\ddot{o}$bius strip universe solution. Again we can show that this solution  describes the accelerated and decelerated  expansion phases of the universe.
\subsubsection{Example 2}
 For the system \eqref{KN5.54}--\eqref{KN5.57} the Mobious  solution reads as
    \begin{eqnarray}
H_1&=&\left(1+\frac{1}{2}\Lambda \cos (\frac{\tau}{2})\right)\cos (\tau), \\
H_2&=&\left(1+\frac{1}{2}\Lambda\cos(\frac{\tau}{2})\right)\sin(\tau),\\
H_3&=&\frac{1}{2}\Lambda\sin (\frac{\tau}{2}).
    \end{eqnarray}
    
    The corresponding EoS takes the form
    \begin{eqnarray}
\rho&=&\frac{D_{24}}{E_{24}},\\
p_1&=&-\frac{D_{25}}{E_{25}},\\
p_2&=&-\frac{D_{26}}{E_{26}},\\
p_3&=&-\frac{D_{27}}{E_{27}},
\end{eqnarray}
where
\begin{eqnarray}
D_{24}&=&[1+\frac{1}{2}\Lambda\cos(\frac{\tau}{2})]^2\cos(\tau)\sin(\tau)+\frac{\Lambda}{2}[1+\frac{\Lambda}{2}\cos(\frac{\tau}{2})][\sin(\tau)+\cos(\tau)]\sin(\frac{\tau}{2})-\Lambda,\\
E_{24}&=&1,\\
D_{25}&=&-\frac{1}{4}\Lambda\sin(\frac{\tau}{2})\sin(\tau)+[1+\frac{1}{2}\Lambda\cos(\frac{\tau}{2})][\cos(\tau)+ \frac{\Lambda}{2}\sin(\tau)\sin(\frac{\tau}{2})]     +\frac{1}{4}\Lambda\cos(\frac{\tau}{2})+\notag\\& &[1+\frac{1}{2}\Lambda\cos(\frac{\tau}{2})]^2\sin^2(\tau)+\frac{\Lambda^2}{4}\sin^2(\frac{\tau}{2})-\Lambda,\\
E_{25}&=&1,\\
D_{26}&=&-\frac{1}{4}\Lambda\sin(\frac{\tau}{2})\cos(\tau)-[1+\frac{1}{2}\Lambda\cos(\frac{\tau}{2})][\sin(\tau)+\frac{\Lambda}{2}\cos(\tau)\sin(\frac{\tau}{2})]+\frac{1}{4}\Lambda\cos(\frac{\tau}{2})+\notag\\& &
[1+\frac{1}{2}\Lambda\cos(\frac{\tau}{2})]^2\cos^2(\tau)+\frac{\Lambda^2}{4}\sin^2(\frac{\tau}{2})-\Lambda,\\
E_{26}&=&1,\\
D_{27}&=&[1+\frac{\Lambda}{2}\cos(\frac{\tau}{2})-\frac{\Lambda}{4}\sin(\frac{\tau}{2})][\cos(\tau)+\sin(\tau)]+\notag\\& &[1+\frac{\Lambda}{2}\cos(\frac{\tau}{2})]^2[1+\cos(\tau)\sin(\tau)]-\Lambda,\\
E_{27}&=&1.
\end{eqnarray}
 The scalar curvature has the form
 \begin{eqnarray}\label{R13}
R&=&(1/2)\Lambda^2(\cos^2(\tau)+\sin^2(\tau)+\cos(\tau)\sin(\tau))\cos^2((1/2)\tau)+\notag\\& &
+(1/2((\cos(\tau)+\sin(\tau))\Lambda\sin((1/2)\tau)+4\cos^2(\tau)+(2+4\sin(\tau))\cos(\tau)+\notag\\& &
+1-2\sin(\tau)+4\sin^2(\tau)))\Lambda\cos((1/2)\tau)+(1/2)\Lambda^2\sin^2((1/2)\tau)+\notag\\& &
+(1/2(\cos(\tau)+\sin(\tau)))\Lambda\sin((1/2)\tau)+\notag\\& &
+2\cos^2(\tau)+(1/2(4+4\sin(\tau)))\cos(\tau)-2\sin(\tau)+2\sin^2(\tau)
\end{eqnarray}
\begin{figure}[h]
	\centering
		\includegraphics[width=0.5 \textwidth]{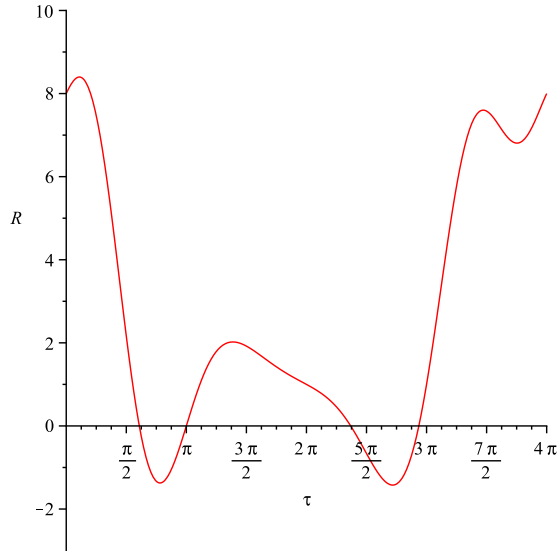}
	\caption{The evolution of the $R$ with respect of the cosmic time $\tau$ for  Eq.\eqref{R13}}
	\label{fig:R5.2.2}
\end{figure}
In Fig.\ref{fig:R5.2.2} we plot the evolution of the $R$ with respect of the cosmic time $\tau$.
\subsection{Other examples of M$\ddot{o}$bius strip like universes induced by Jacobian elliptic functions }
\subsubsection{Example 1} Now we want present some solutions in terms of the Jacobian elliptic functions. In fact, the system \eqref{KN5.50}--\eqref{KN5.53} has the following particular  solution
\begin{eqnarray}
A&=&A_0+\left(1+\frac{1}{2}\Lambda \mbox{cn} \frac{\tau}{2}\right)\mbox{cn} \tau, \\
B&=&B_0+\left(1+\frac{1}{2}\Lambda\mbox{cn}\frac{\tau}{2}\right)\mbox{sn}\tau,\\
C&=&C_0+\frac{1}{2}\Lambda\mbox{sn} \frac{\tau}{2}. 
    \end{eqnarray}
    
\begin{figure}[h]
	\centering
		\includegraphics[width=0.5 \textwidth]{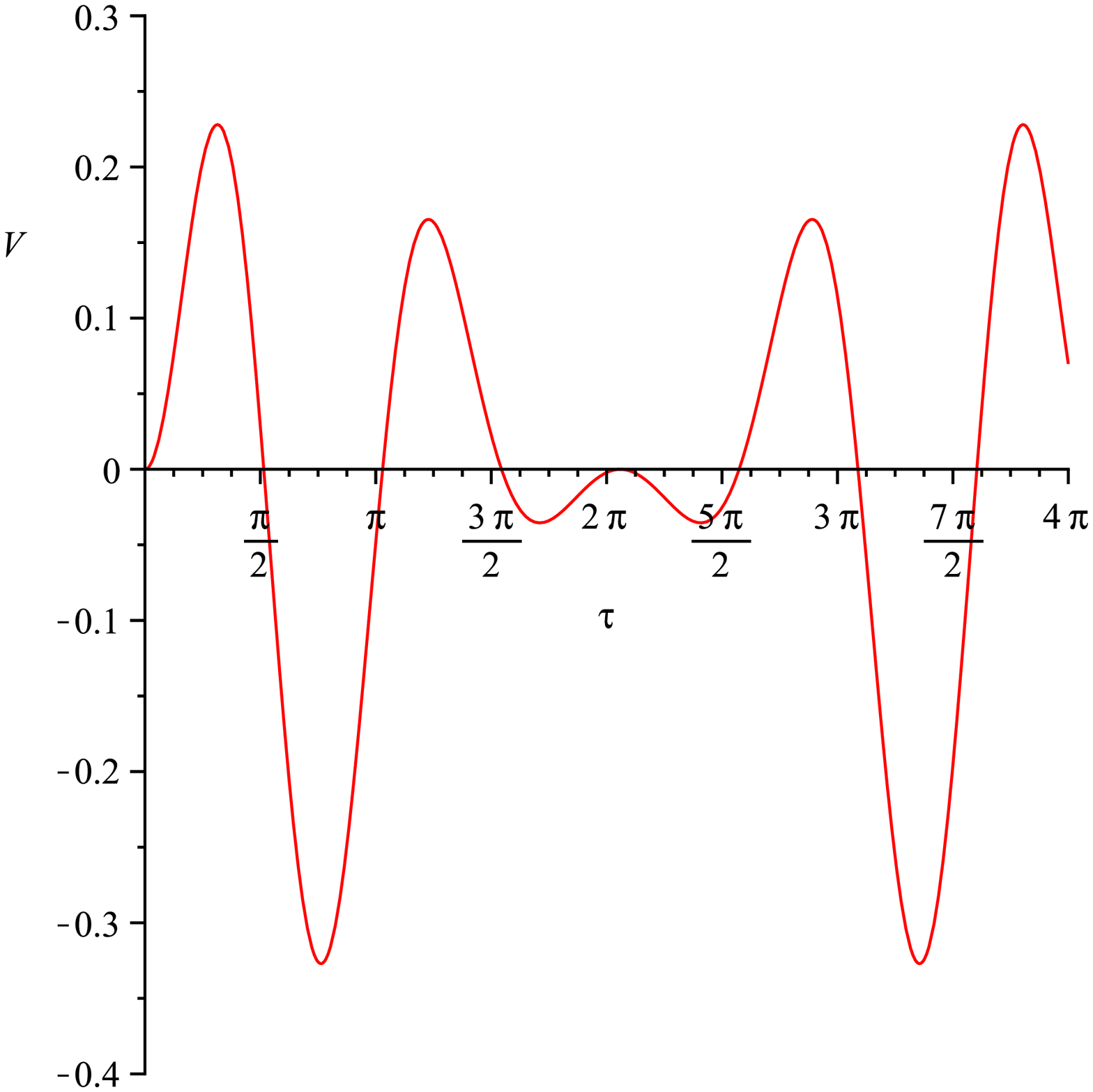}
	\caption{The  evolution of the volume of the trefoil knot universe with respect to the cosmic time $\tau$ for  Eq.\eqref{V14}}
	\label{fig:V5.3.1}
\end{figure}

    The corresponding EoS takes the form
    \begin{eqnarray}
\rho&=&\frac{D_{28}}{E_{28}},\\
p_1&=&-\frac{D_{29}}{E_{29}},\\
p_2&=&-\frac{D_{30}}{E_{30}},\\
p_3&=&-\frac{D_{31}}{E_{31}},
\end{eqnarray}
where
\begin{eqnarray}
D_{28}&=&[\frac{1}{4}\Lambda \mbox{dn}\frac{\tau}{2}\  \mbox{sn}\frac{\tau}{2}\ \mbox{cn}\tau +(1+\frac{1}{2}\Lambda \mbox{cn}\frac{\tau}{2})\ \mbox{dn}\tau\ \mbox{sn}\tau] [\frac{1}{4}\Lambda\mbox{dn}\frac{\tau}{2}\ \mbox{sn}\frac{\tau}{2}\ \mbox{sn}\tau-(1+\frac{1}{2}\Lambda\mbox{cn}\frac{\tau}{2})\ \mbox{cn}\tau\ \mbox{dn}\tau]\times\notag\\& & [C_0+\frac{1}{2}\Lambda\mbox{sn}\frac{\tau}{2}]+\frac{\Lambda}{4} [-\frac{1}{4}\Lambda\mbox{dn}\frac{\tau}{2}\ \mbox{sn}\frac{\tau}{2}\ \mbox{sn}\tau+(1+\frac{1}{2}\Lambda\mbox{cn}\frac{\tau}{2})\ \mbox{cn}\tau\ \mbox{dn}\tau]\mbox{cn}\frac{\tau}{2}\ \mbox{dn}\frac{\tau}{2}\times\notag\\& &
[A_0+(1+\frac{1}{2}\Lambda\mbox{cn}\frac{\tau}{2})\mbox{cn}\tau]-\frac{1}{4} [\frac{1}{4}\Lambda\mbox{dn}\frac{\tau}{2}\ \mbox{sn}\frac{\tau}{2}\ \mbox{cn}\tau
+(1+\frac{1}{2}\Lambda\mbox{cn}\frac{\tau}{2})\ \mbox{dn}\tau\ \mbox{sn}\tau]\Lambda\mbox{cn}\frac{\tau}{2}\ \mbox{dn}\frac{\tau}{2}\times\notag\\& & [B_0+(1+\frac{1}{2}\Lambda\mbox{cn}\frac{\tau}{2})\ \mbox{sn}\tau]-\Lambda [A_0+(1+\frac{1}{2}\Lambda\mbox{cn}\frac{\tau}{2})\ \mbox{cn}\tau] [B_0+(1+\frac{1}{2}\Lambda\mbox{cn}\frac{\tau}{2})\ \mbox{sn}\tau]\times\notag\\& & [C_0+\frac{1}{2}\Lambda\mbox{sn}\frac{\tau}{2}],\\
E_{28}&=&[A_0+(1+\frac{1}{2}\Lambda\mbox{cn}\frac{\tau}{2})\ \mbox{cn}\tau] [B_0+(1+\frac{1}{2}\Lambda\mbox{cn}\frac{\tau}{2})\ \mbox{sn}\tau] [C_0+\frac{1}{2}\Lambda\mbox{sn}\frac{\tau}{2}],\\
D_{29}&=&[C_0+\frac{1}{2}\Lambda\mbox{sn}\frac{\tau}{2}] [\frac{1}{8}\Lambda \mbox{cn}\frac{\tau}{2}\ \mbox{sn}^2\frac{\tau}{2}\ \mbox{sn}\tau-\frac{1}{8}\Lambda\mbox{dn}^2\frac{\tau}{2}\ \mbox{cn}\frac{\tau}{2}\ \mbox{sn}\tau-\frac{1}{2}\Lambda\mbox{dn}\frac{\tau}{2}\ \mbox{sn}\frac{\tau}{2}\ \mbox{cn}\tau\ \mbox{dn}\tau-\notag\\& & (1+\frac{1}{2}\Lambda\mbox{cn}\frac{\tau}{2})\ \mbox{dn}^2\tau\ \mbox{sn}\tau-[1+\frac{1}{2}\Lambda\mbox{cn}\frac{\tau}{2}]\ \mbox{cn}^2\tau \ \mbox{sn}\tau]+[B_0+(1+\frac{1}{2}\Lambda\mbox{cn}\frac{\tau}{2})\ \mbox{sn}\tau]\times\notag\\& & [-\frac{1}{8}\Lambda\mbox{dn}^2\frac{\tau}{2}\ \mbox{sn}\frac{\tau}{2}-\frac{1}{8}\Lambda\mbox{cn}^2\frac{\tau}{2} \mbox{sn}\frac{\tau}{2}]-\frac{\Lambda}{4} [\frac{1}{4}\Lambda\mbox{dn}\frac{\tau}{2}\ \mbox{sn}\frac{\tau}{2}\ \mbox{sn}\tau-(1+\frac{1}{2}\Lambda\mbox{cn}\frac{\tau}{2})\ \mbox{cn}\tau\ \mbox{dn}\tau]\times\notag\\& & \mbox{cn}\frac{\tau}{2}\ \mbox{dn}\frac{\tau}{2}-\Lambda [B_0+(1+\frac{1}{2}\Lambda\mbox{cn}\frac{\tau}{2})\ \mbox{sn}\tau] [C_0+\frac{1}{2}\Lambda\mbox{sn}\frac{\tau}{2}],\\
E_{29}&=&[B_0+(1+\frac{1}{2}\Lambda\mbox{cn}\frac{\tau}{2})\ \mbox{sn}\tau] [C_0+\frac{1}{2}\Lambda\mbox{sn}\frac{\tau}{2}],\\
D_{30}&=&-[A_0+(1+\frac{1}{2}\Lambda\ \mbox{cn}\frac{\tau}{2})\ \mbox{cn}\tau] [\frac{1}{8}\Lambda\ \mbox{dn}^2\frac{\tau}{2}\ \mbox{sn}\frac{\tau}{2}+\frac{1}{8}\Lambda\ \mbox{cn}^2\frac{\tau}{2}\ \mbox{sn}\frac{\tau}{2}]+[C_0+\frac{1}{2}\Lambda\ \mbox{sn}\frac{\tau}{2}]\times\notag\\& &
 [\frac{1}{8}\Lambda \ \mbox{cn}\frac{\tau}{2}\ \mbox{sn}^2\frac{\tau}{2}\ \mbox{cn}\tau-\frac{1}{8}\Lambda\ \mbox{dn}^2\frac{\tau}{2}\ \mbox{cn}\frac{\tau}{2}\ \mbox{cn}\tau+\frac{1}{2}\Lambda\ \mbox{dn}\frac{\tau}{2}\ \mbox{sn}\frac{\tau}{2}\ \mbox{dn}\tau\ \mbox{sn}\tau+\notag\\& &
 (1+\frac{1}{2}\Lambda\ \mbox{cn}\frac{\tau}{2})(\mbox{cn}\tau\ \mbox{sn}^2\tau- \mbox{dn}^2\tau\ \mbox{cn}\tau)]-\frac{\Lambda}{4} [\frac{1}{4}\Lambda\ \mbox{dn}\frac{\tau}{2}\ \mbox{sn}\frac{\tau}{2}\ \mbox{cn}\tau+(1+\frac{1}{2}\Lambda\ \mbox{cn}\frac{\tau}{2})\times\notag\\& & \mbox{dn}\tau\ \mbox{sn}\tau]\ \mbox{cn}\frac{\tau}{2}\ \mbox{dn}\frac{\tau}{2}-\Lambda [A_0+(1+\frac{1}{2}\Lambda\ \mbox{cn}\frac{\tau}{2})\ \mbox{cn}\tau] [C_0+\frac{1}{2}\Lambda\ \mbox{sn}\frac{\tau}{2}],\\
E_{30}&=&[A_0+(1+\frac{1}{2}\Lambda\ \mbox{cn}\frac{\tau}{2})\ \mbox{cn}\tau] [C_0+\frac{1}{2}\Lambda\ \mbox{sn}\frac{\tau}{2}],\\
D_{31}&=&[B_0+(1+\frac{1}{2}\Lambda\ \mbox{cn}\frac{\tau}{2})\ \mbox{sn}\tau] [\frac{1}{8}\Lambda\ \mbox{cn}\frac{\tau}{2}\ \mbox{sn}^2\frac{\tau}{2}\ \mbox{cn}\tau-\frac{1}{8}\Lambda\ \mbox{dn}^2\frac{\tau}{2}\ \mbox{cn}\frac{\tau}{2}\ \mbox{cn}\tau+\notag\\& & 
\frac{1}{2}\Lambda\ \mbox{dn}\frac{\tau}{2}\ \mbox{sn}\frac{\tau}{2}\ \mbox{dn}\tau\ \mbox{sn}\tau+(1+\frac{1}{2}\Lambda\ \mbox{cn}\frac{\tau}{2}) (\mbox{sn}^2\tau-\ \mbox{dn}^2\tau)\ \mbox{cn}\tau]+
[A_0+(1+\frac{1}{2}\Lambda\ \mbox{cn}\frac{\tau}{2})\ \mbox{cn}\tau]\times\notag\\& & 
[\frac{1}{8}\Lambda \ \mbox{cn}\frac{\tau}{2}\  \mbox{sn}\tau(\mbox{sn}^2\frac{\tau}{2}-\mbox{dn}^2\tau)-\frac{1}{2}\Lambda\ \mbox{dn}\frac{\tau}{2}\ \mbox{sn}\frac{\tau}{2}\ \mbox{cn}\tau\ \mbox{dn}\tau-(1+\frac{1}{2}\Lambda\ \mbox{cn}\frac{\tau}{2})\ \mbox{dn}^2\tau\ \mbox{sn}\tau-\notag\\& &
(1+\frac{1}{2}\Lambda\ \mbox{cn}\frac{\tau}{2})\ \mbox{cn}^2\tau \ \mbox{sn}\tau]+[\frac{1}{4}\Lambda\ \mbox{dn}\frac{\tau}{2}\ \mbox{sn}\frac{\tau}{2}\ \mbox{cn}\tau+(1+\frac{1}{2}\Lambda\ \mbox{cn}\frac{\tau}{2})\ \mbox{dn}\tau\ \mbox{sn}\tau]\times\notag\\& &
 [\frac{1}{4}\Lambda\ \mbox{dn}\frac{\tau}{2}\ \mbox{sn}\frac{\tau}{2}\ \mbox{sn}\tau-(1+\frac{1}{2}\Lambda\ \mbox{cn}\frac{\tau}{2})\ \mbox{cn}\tau\ \mbox{dn}\tau]-\Lambda [A_0+(1+\frac{1}{2}\Lambda\ \mbox{cn}\frac{\tau}{2})\ \mbox{cn}\tau]\times\notag\\& & [B_0+(1+\frac{1}{2}\Lambda\ \mbox{cn}\frac{\tau}{2})\ \mbox{sn}\tau],\\
E_{31}&=&[B_0+(1+\frac{1}{2}\Lambda\ \mbox{cn}\frac{\tau}{2})\ \mbox{sn}\tau] [A_0+(1+\frac{1}{2}\Lambda\ \mbox{cn}\frac{\tau}{2})\ \mbox{cn}\tau].
\end{eqnarray}
The evolution of the volume of the universe for \eqref{KN3.16} reads as ($A_0=B_0=C_0=0$)
  \begin{equation}\label{V14}
V=\frac{1}{2}\Lambda\left(1+\frac{1}{2}\Lambda \mbox{cn} \frac{\tau}{2}\right)^2\mbox{cn} \tau\ \mbox{sn}\tau\ \mbox{sn} \frac{\tau}{2}.
\end{equation}
The  evolution of the volume with \eqref{KN3.16} and $\Lambda =1$  presented in Fig.\ref{fig:V5.3.1}.

 The scalar curvature has the form
 \begin{eqnarray}\label{R14}
R&=&(2\Lambda(2k^2\mbox{cn}((1/2)\tau, k)+\Lambda k^2\mbox{cn}^2((1/2)\tau, k)+\notag\\& &
+(1/2)\Lambda\mbox{dn}^2((1/2)\tau, k))\mbox{cn}(\tau, k)\mbox{sn}(\tau, k)\mbox{sn}((1/2)\tau, k)^3-\notag\\& &
-6\Lambda\mbox{dn}(\tau, k)\mbox{dn}((1/2)\tau, k)(\mbox{cn}(\tau, k)-\mbox{sn}(\tau, k))(\mbox{cn}(\tau, k)+\notag\\& &
+\mbox{sn}(\tau, k))(2+\Lambda\mbox{cn}((1/2)\tau, k))\mbox{sn}^2((1/2)\tau, k)-\notag\\& &
-(4(2+\Lambda\mbox{cn}((1/2)\tau, k)))((1/4)\mbox{cn}^3((1/2)\tau, k)k^2\Lambda+\notag\\& &
+(1/2)\mbox{cn}^2((1/2)\tau, k)k^2+\Lambda(3\mbox{dn}^2(\tau, k)+\mbox{cn}^2(\tau, k)k^2+(5/4)\mbox{dn}^2((1/2)\tau, k)-\notag\\& &
-\mbox{sn}^2(\tau, k)k^2)\mbox{cn}((1/2)\tau, k)+6\mbox{dn}^2(\tau, k)+(1/2)\mbox{dn}^2((1/2)\tau, k)-\notag\\& &
-2\mbox{sn}^2(\tau, k)k^2+2\mbox{cn}^2(\tau, k)k^2)\mbox{cn}(\tau, k)\mbox{sn}(\tau, k)\mbox{sn}((1/2)\tau, k)+\notag\\& &
+2\mbox{cn}((1/2)\tau, k)\mbox{dn}((1/2)\tau, k)\mbox{dn}(\tau, k)(\mbox{cn}(\tau, k)-\mbox{sn}(\tau, k))(\mbox{cn}(\tau, k)+\notag\\& &
+\mbox{sn}(\tau, k))(2+\Lambda\mbox{cn}((1/2)\tau, k))^2)/(\mbox{cn}(\tau, k)\mbox{sn}(\tau, k)*\notag\\& &
*(2+\Lambda\mbox{cn}^2((1/2)\tau, k))\mbox{sn}((1/2)\tau, k)).
\end{eqnarray}
\begin{figure}[h]
	\centering
		\includegraphics[width=0.5 \textwidth]{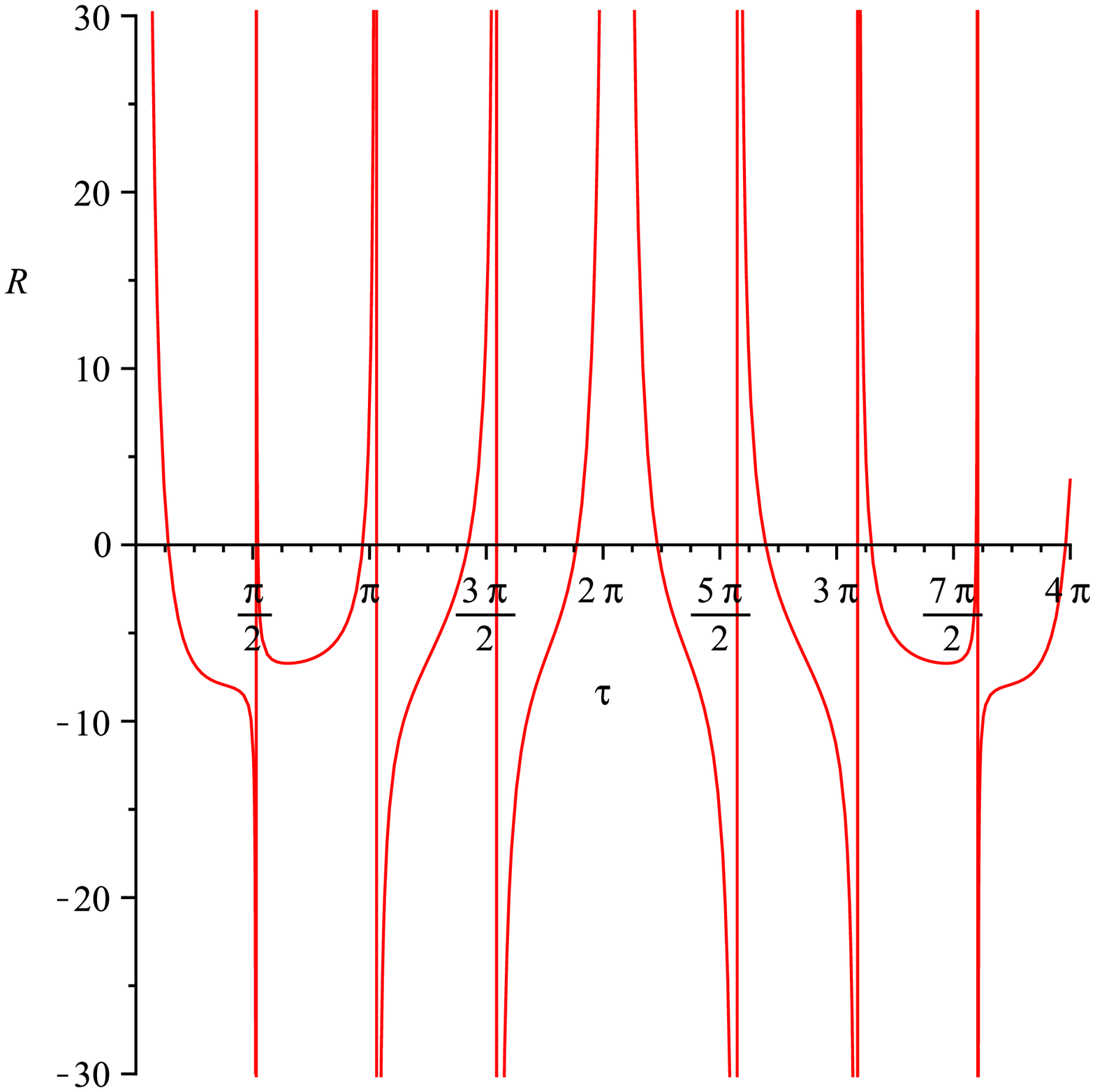}
	\caption{The evolution of the $R$ with respect of the cosmic time $\tau$ for  Eq.\eqref{R14}}
	\label{fig:R5.3.1}
\end{figure}
In Fig.\ref{fig:R5.3.1} we plot the evolution of the $R$ with respect of the cosmic time $\tau$.
\subsubsection{Example 2}   Similarly, we can show that the system \eqref{KN5.54}--\eqref{KN5.57} has the following solution
\begin{eqnarray}
H_1&=&\left(1+\frac{1}{2}\Lambda \mbox{cn} \frac{\tau}{2}\right)\mbox{cn} \tau, \\
H_2&=&\left(1+\frac{1}{2}\Lambda\mbox{cn}\frac{\tau}{2}\right)\mbox{sn}\tau,\\
H_3&=&\frac{1}{2}\Lambda\mbox{sn} \frac{\tau}{2}.
    \end{eqnarray}
    
    The corresponding EoS takes the form
    \begin{eqnarray}
\rho&=&\frac{D_{32}}{E_{32}},\\
p_1&=&-\frac{D_{33}}{E_{33}},\\
p_2&=&-\frac{D_{34}}{E_{34}},\\
p_3&=&-\frac{D_{35}}{E_{35}},
\end{eqnarray}
where
\begin{eqnarray}
D_{32}&=&[1+\frac{1}{2}\ \Lambda \ \mbox{cn}\frac{\tau}{2}]^2 \ \mbox{cn}\tau \ \mbox{sn}\tau+\frac{\Lambda}{2}\ [1+\frac{1}{2} \Lambda \ \mbox{cn}\frac{\tau}{2}][\mbox{sn}\tau+\mbox{cn}\tau] \ \mbox{sn}\frac{\tau}{2}-\Lambda,\\
E_{32}&=&1,\\
D_{33}&=&\frac{1}{4}\Lambda\mbox{dn}\frac{\tau}{2}\  \mbox{sn}\frac{\tau}{2}[1-\mbox{sn}\tau]+[1+\frac{1}{2} \Lambda\ \mbox{cn}\frac{\tau}{2}] [\mbox{cn}\tau \ \mbox{dn}\tau+\frac{\Lambda}{2}\  \mbox{sn}\tau\  \mbox{sn}\frac{\tau}{2}]+[1+\frac{1}{2} \Lambda  \mbox{cn}\frac{\tau}{2}]^2\  \mbox{sn}^2\tau+\notag\\& &
\frac{1}{4}\Lambda^2\mbox{sn}^2\frac{\tau}{2}-\Lambda,\\
E_{33}&=&1,\\
D_{34}&=&-\frac{1}{4} \Lambda \ \mbox{dn}\frac{\tau}{2} \ \mbox{sn}\frac{\tau}{2}\  \mbox{cn}\tau-(1+\frac{1}{2} \Lambda \ \mbox{cn}\frac{\tau}{2})[\mbox{dn}\tau\  \mbox{sn}\tau+\frac{\Lambda}{2}\  \mbox{cn}\tau  \ \mbox{sn}\frac{\tau}{2}]+\frac{1}{4} \Lambda\  \mbox{cn}\frac{\tau}{2} \ \mbox{dn}\frac{\tau}{2}+\notag\\& &[1+\frac{1}{2} \Lambda\  \mbox{cn}\frac{\tau}{2}]^2 \ \mbox{cn}^2\tau+
\frac{1}{4}\ \Lambda^2\  \mbox{sn}^2\frac{\tau}{2}-\Lambda,\\
E_{34}&=&1,\\
D_{35}&=&-\frac{1}{4}\Lambda\  \mbox{dn}\frac{\tau}{2}\  \mbox{sn}\frac{\tau}{2}[\mbox{cn}\tau+\mbox{sn}\tau]+[1+\frac{1}{2} \Lambda \ \mbox{cn}\frac{\tau}{2}][\mbox{cn}\tau-\mbox{sn}\tau]  \mbox{dn}\tau+\notag\\& &[1+\frac{1}{2}\ \Lambda\  \mbox{cn}\frac{\tau}{2}]^2[\mbox{sn}^2\tau+  \mbox{cn}^2\tau+  \mbox{cn}\tau\  \mbox{sn}\tau]-\Lambda,\\
E_{35}&=&1.
\end{eqnarray}
   
 The scalar curvature has the form
 \begin{eqnarray}\label{R15}
R&=&(1/2)\Lambda^2(\mbox{cn}^2(\tau, k)+\mbox{sn}^2(\tau, k)+\mbox{cn}(\tau, k)\mbox{sn}(\tau, k))\mbox{cn}^2((1/2)\tau, k)+\notag\\& &
+(1/2(\Lambda(\mbox{sn}(\tau, k)+\mbox{cn}(\tau, k))\mbox{sn}((1/2)\tau, k)+\mbox{dn}((1/2)\tau, k)+4\mbox{cn}^2(\tau, k)+\notag\\& &
+(4\mbox{sn}(\tau, k)+2\mbox{dn}(\tau, k))\mbox{cn}(\tau, k)+4\mbox{sn}^2(\tau, k)-\notag\\& &
-2\mbox{dn}(\tau, k)\mbox{sn}(\tau, k)))\Lambda\mbox{cn}((1/2)\tau, k)+(1/2)\Lambda^2\mbox{sn}^2((1/2)\tau, k)-\notag\\& & 
-(1/2)\Lambda(\mbox{dn}((1/2)\tau, k)-2)(\mbox{sn}(\tau, k)+\notag\\& &
+\mbox{cn}(\tau, k))\mbox{sn}((1/2)\tau, k)+2\mbox{cn}^2(\tau, k)+(1/2(4\mbox{dn}(\tau, k)+\notag\\& &
+4\mbox{sn}(\tau, k)))\mbox{cn}(\tau, k)-2\mbox{sn}(\tau, k)(-\mbox{sn}(\tau, k)+\mbox{dn}(\tau, k)).
\end{eqnarray}
\begin{figure}[h]
	\centering
		\includegraphics[width=0.5 \textwidth]{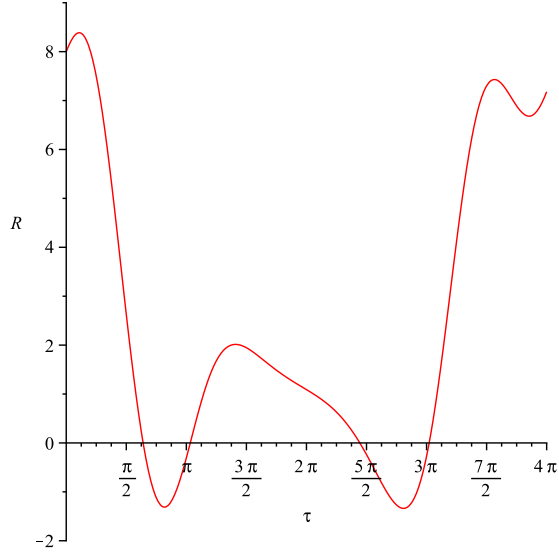}
\caption{The evolution of the $R$ with respect of the cosmic time $\tau$ for  Eq.\eqref{R15}}
\label{fig:R5.3.2}
\end{figure}
In Fig.\ref{fig:R5.3.2} we plot the evolution of the $R$ with respect of the cosmic time $\tau$.

%%%%%%%%%%%%%%%%%%%%%%%%%%%%%%%%%%%%%%%%%%%%%%
\subsection{Integrable models}
%%%%%%%%%%%%%%%%%%%%%%%%%%%%%%%%%%%%%%%%%%%%%
The system (2.21)-(2.24) contents 4 equations for 7 unknown functions. This means we can add 3 new additional equations. It gives us in particular to construct integrable Bianchi models. In this subsection we present two examples of such integrable models.
%%%%%%%%%%%%%%%%%%%%%%%%%%%%%%%%%%%%%%%%%%%%%%
\subsubsection{Euler top equation}
%%%%%%%%%%%%%%%%%%%%%%%%%%%%%%%%%%%%%%%%%%%%%
Let us we assume that $A=A_1, B=A_2, C=A_3$ obey the Euler top equation. 
The simple Euler top equation reads as
\begin{eqnarray}\label{R15}
\dot{A}_{1}&=&A_2A_3, \\
\dot{A}_{2}&=&A_3A_1, \\
\dot{A}_{3}&=&A_1A_2.
\end{eqnarray}
The system of equations (2.21)-(2.24) and (5.124)-(5.126) we call the  Bianchi I-Euler model which is integrable.

%%%%%%%%%%%%%%%%%%%%%%%%%%%%%%%%%%%%%%%%%%%%%%
\subsubsection{Heisenberg ferromagnet equation}
%%%%%%%%%%%%%%%%%%%%%%%%%%%%%%%%%%%%%%%%%%%%%
Our second example is the  Heisenberg ferromagnet equation (HFE). Here we assume that the variables $A=S_1, B=S_2, C=S_3$ satisfy the equation
$iS_{t}+\frac{1}{\omega}[S, W]=0,\quad  iW_{x}+\omega [S, W]=0,$ where $S=S_i\sigma_i$, $W=W_i\sigma_i$ and  $\sigma_i$ are Pauli's matrices.  
 It  the principal chiral type equation with the $U=-i\lambda S,\quad V=-\frac{i\lambda}{\omega(\lambda+\omega)}W.$. Note this principal chiral type equation
 is the particular case of the following  (2+1)-dimensional M-XCIX equation \cite{M01} (see also \cite{M02})
 \begin{eqnarray}
iS_{t}+0.5([S, S_{y}]+uS)_{x}+\frac{1}{\omega}[S, W]&=&0,\\
u_x-0.5S\cdot[S_x,S_y]&=&0,\\
 iW_{x}+\omega [S, W]&=&0 \end{eqnarray} 
 or (that equivalent) 
 \begin{eqnarray}
iS_{t}+0.5[S, S_{xy}]+uS_{x}+\frac{1}{\omega}[S, W]&=&0,\\
u_x-0.5S\cdot[S_x,S_y]&=&0,\\
 iW_{x}+\omega [S, W]&=&0. \end{eqnarray} 
 This M-XCIX equation is integrable by the following  Lax representation \cite{M01}
  \begin{eqnarray}
\Phi_{x}&=&U\Phi,\\
\Phi_{t}&=&2\lambda\Phi_y+V\Phi, 
\end{eqnarray}  
with ($S^2=I$)
 \begin{eqnarray}
U&=&-i\lambda S, \\
V&=&\lambda V_{1}+\frac{i}{\lambda+\omega}W-\frac{i}{\omega}W 
\end{eqnarray} 
and 
\begin{eqnarray}
S&=&\begin{pmatrix} S_3&S^{-}\\S^{+}& -S_3\end{pmatrix},\\
V_1&=&0.25([S, S_{y}]+uS),\\
W&=&\begin{pmatrix} W_3&W^{-}\\W^{+}& -W_3\end{pmatrix}. 
\end{eqnarray}

%%%%%%%%%%%%%%%%%%%%%%%%%%%%%%%%%%%%%%%%%%%%%%
\section{Conclusion}
%%%%%%%%%%%%%%%%%%%%%%%%%%%%%%%%%%%%%%%%%%%%%

In the present paper, we have constructed several concrete models describing the trefoil and figure-eight knot universes from Bianchi-type I cosmology and 
examined the cosmological features and properties in detail. 

To realize the cyclic universes, 
it is necessary to a non-canonical scalar field with non well-defined vacuum 
in the context of the quantum field theory 
or extend gravity, e.g., with adding higher order derivative terms 
and $f(R)$ gravity~\cite{Cai:2011bs}. Indeed, however, 
these modified gravity theories have to satisfy the tests on the solar system scale as well as cosmological constraints so that those can be alternative gravitational theories to general relativity. 
The significant cosmological consequence of our approach is that we have 
shown the possibility to obtain the knot universes related to the cyclic 
universes from Bianchi-type I spacetime within general relativity. 

Furthermore, recently it has been pointed out that 
the asymmetry of the EoS for the universe can 
lead to cosmological hysteresis~\cite{Sahni:2012er}. 
On the other hand, 
Bianchi-type I spacetime describes the spatially anisotropic cosmology and hence the EoS for the universe has the asymmetry 
in the oscillating process through the expanding and contracting behaviors. 
Accordingly, it is considered that in the constructed models of the knot universes cosmological hysteresis could occur. The observation of this phenomenon in our models is one of our future works on the knot universes. 

Finally, it should be remarked that 
by summarizing the results of our previous~\cite{Knot-universe}
and this works, the knot universes describing the cyclic universes 
can be realized from the homogeneous and isotropic FLRW spacetime as well as 
the homogeneous and anisotropic Bianchi-type I cosmology. In these series of works, the formulations of model construction method of the knot universes have been established. 
Thus, it can be expected that the presented formalism is useful to 
realize the universes with other features from both 
the isotropic and anisotropic spacetimes. 

Finally we would like to note that all solutions presented above   describe the accelerated and decelerated  expansion phases of the universe.

  \end{document}